\documentclass[final,3p,times]{elsarticle}  %
\biboptions{sort&compress}

\usepackage{enumerate}
\usepackage{amsmath,amssymb}
\usepackage{graphicx}
\usepackage{slashed}
\usepackage{float}
\usepackage[dvipsnames]{xcolor}
\usepackage[normalem]{ulem}
\usepackage{subfigure,orcidlink}
\usepackage{multirow,array}
\usepackage{adjustbox}
\usepackage{hyperref}
\hypersetup{%
	colorlinks = true,%
	linkcolor = Blue,%
	citecolor = Blue,%
	filecolor = Blue,%
	urlcolor = Blue%
}
\usepackage{tabularray}
\UseTblrLibrary{booktabs}
\usepackage{tabularx}

\journal{Nuclear Physics B}
\makeatletter
\def\ps@pprintTitle{%
  \let\@oddhead\@empty
  \let\@evenhead\@empty
  \def\@oddfoot{\footnotesize\itshape \href{https://www.sciencedirect.com/science/article/pii/S0550321326001859?via\%3Dihub}{Nucl. Phys. B 1027 (2026) 117478} \hfill \today}%
  \let\@evenfoot\@oddfoot}
\makeatother

\usepackage{fancyhdr}
\begin{document}
\begin{frontmatter}

\title{Cosmological Probes of Lepton Parity Freeze-in Dark Matter:\\$\Delta N_{\rm eff}$ \& Gravitational Waves}

\cortext[cor1]{Corresponding author}

\author[ucr]{Ernest Ma}

\address[ucr]{Department of Physics and Astronomy,
University of California, Riverside, California 92521, USA}

\author[iith]{Partha Kumar Paul\orcidlink{0000-0002-9107-5635}\corref{cor1}}
\ead{ph22resch11012@iith.ac.in}

\author[iith]{Narendra Sahu\orcidlink{0000-0002-9675-0484}}
\ead{nsahu@phy.iith.ac.in}

\address[iith]{Department of Physics, Indian Institute of Technology Hyderabad, Kandi, Telangana-502285, India}

\date{\today}


\begin{abstract}
In the canonical type-I seesaw mechanism for neutrino masses, a residual symmetry known as lepton parity: $(-1)^L$, remains preserved. Introducing a Majorana fermion $S$ with even lepton parity renders it naturally stable, making it a viable dark matter (DM) candidate. The addition of a lepton parity odd singlet scalar $\sigma$ allows for the coupling $N S \sigma$, where $N$ is the right-handed neutrino. If $S$ is not thermalized, then DM relic can be produced in two distinct ways: (i) for reheating temperature, $T_{\rm rh}>m_{N}$, dominantly through the decay of $N$ ($N\rightarrow S\sigma$), and (ii) for $T_{\rm EW}<T_{\rm rh}\ll m_{N}$, via standard model Higgs ($h$) decay ($h\rightarrow SS$ at one loop). If the $\sigma-h$ quartic coupling is large, then it can lead to a strong first-order electroweak phase transition even if $\langle\sigma\rangle=0$.  Alternatively, if $\sigma-h$ coupling is small, then $\sigma$ can freeze out with a larger abundance, and hence its decay ($\sigma\rightarrow S\nu$) at late epochs can give rise to additional relativistic degrees of freedom ($\Delta{N}_{\rm eff}$). Thus, the framework gives a viable DM with mass range varying from MeV to TeV and leaves observable imprints, via gravitational waves and $\Delta{N}_{\rm eff}$, which offer complementary probes, potentially detectable in future gravitational wave and CMB experiments.
\end{abstract}


\end{frontmatter}

\section{Introduction}\label{sec:intro}

To accommodate naturally small Majorana neutrino masses, the standard model (SM) of quarks  
and leptons is routinely extended to include three singlet right-handed neutrinos (RHN) $N$ 
with large Majorana masses, implementing thus the well-known type-I seesaw mechanism \cite{Minkowski:1977sc, Gell-Mann:1979vob, Mohapatra:1979ia, Schechter:1980gr}. 
The resulting model conserves lepton parity $(-1)^L$, which may be used to define 
dark parity $(-1)^{(L+2j)}$ \cite{Ma:2015xla}.  Suppose a singlet left-handed Majorana fermion $S$ is added with even lepton parity (hence odd dark parity), then it is a dark matter (DM) 
candidate.  To connect it with the SM, a singlet real scalar $\sigma$ may be added. 
The case of $\sigma$ even under lepton(dark) parity has been discussed in two recent 
papers \cite{Ma:2024dhk, Ma:2025bjf}. The former considers the case where $m^2_\sigma > 0$ \cite{Ma:2024dhk}, resulting in a small vacuum expectation value 
(VEV) of $\sigma$ and thereby giving a possibility of light fermion DM. The latter considers the case where $m^2_\sigma < 0$ \cite{Ma:2025bjf}, resulting in a large VEV of $\sigma$ leading to the formation of unstable domain walls whose disappearance gives gravitational waves. Here we consider 
the other possible option, \textit{i.e.}, $\sigma$ is odd under lepton (dark) parity with no VEV, and study the related phenomenologies beyond the SM. Unlike the previous studies, in the present case, the charge assignment of $\sigma$ permits a new Yukawa coupling $y_{NS} N S \sigma$, which opens the direct freeze-in channel $N_1 \to S\sigma$ and simultaneously enables the late decay $\sigma \to S\nu$ via $N-\nu$ mixing. These typical features were absent in the previous realizations. Moreover, since $\langle \sigma \rangle = 0$, the stability of $S$ is automatic without additional assumptions. Additionally, the presence of $\sigma$ in this setup gives rise to a strong electroweak first-order phase transition \cite{Curtin:2014jma,Dey:2025pcs} in the early Universe, which results in stochastic gravitational waves. This can give rise to detectable signatures at different gravitational wave experiments.

When $\sigma^2H^\dagger H$ coupling is large, $\sigma$ will freeze out with a smaller abundance. In this scenario, the late-time decay of $\sigma$ gives a negligible contribution to the DM relic. If DM $S$ is not thermalized, then its relic can be obtained from $N$ or Higgs decay, depending on the reheat temperature of the Universe. Alternatively, if $\sigma^2H^\dagger H$ coupling is small, then $\sigma$ can freeze out with a larger abundance in the early Universe. The late decay of $\sigma$ through $\sigma\rightarrow S\nu$ channel can give rise to the non-thermal production of light degrees of freedom, ${N}_{\rm eff}$, as well as DM. In this scenario, if we demand that the DM relic from $\sigma$ decay is in the correct range, then it results in a larger $\Delta{N}_{\rm eff}$ which is ruled out by the present CMB experiments. On the other hand, if we demand the $\sigma$ decay produces the observable $\Delta{N}_{\rm eff}$, then the relic of $S$ remains underabundant. The deficit can be fulfilled through $N$ or $H$ decay, depending on the reheating temperature of the Universe. Thus, it gives rise to a complementary probe of the scenario at different cosmic microwave background (CMB) experiments such as  DESI \cite{DESI:2024mwx}, Planck \cite{Planck:2018vyg}, ACT \cite{ACT:2025tim}, Planck+ACT \cite{ACT:2025tim}, SPT-3G \cite{SPT-3G:2019sok}, CMB-S4 \cite{Abazajian:2019eic}, and CMB-HD \cite{CMB-HD:2022bsz}.

The paper is organized as follows. In section \ref{sec:model}, we discuss the minimal model in detail. Section \ref{sec:sddmrelic} is devoted to discussing the DM phenomenology and its probe at terrestrial experiments when the reheat temperature is large ($T_{\rm rh}>m_{N}$). The DM phenomenology with low reheat temperature ($T_{\rm EW}<T_{\rm rh}\ll m_{N}$) is discussed in section \ref{sec:lowTrh}. We finally conclude in section \ref{sec:concl}.

\section{Minimal model}\label{sec:model}

In the minimal type-I seesaw framework, lepton number is explicitly broken down to a residual lepton parity symmetry $(-1)^L$. By construction, the SM leptons are odd under this residual lepton parity, and we choose the RHN, $N$, to be also odd under lepton parity. In this work, we extend the minimal type-I seesaw model by introducing a left-handed fermion, $S$, which carries even lepton parity and is therefore disconnected from the SM sector. This makes $S$ a natural DM candidate. To enable its production in the early Universe, we further introduce a singlet scalar, $\sigma$, which is odd under lepton parity with a mass hierarchy as $m_N>m_\sigma>m_S$. The relevant Lagrangian is then given as
\begin{eqnarray}
    \mathcal{L}=\overline{N}i\gamma^\mu\partial_\mu N-\frac{1}{2}m_{N_i}\overline{N_i^C}N_i+\overline{S}i\gamma^\mu\partial_\mu S-\frac{1}{2}m_S\overline{S^C}S-y_{\alpha i}\bar{L}_\alpha\tilde{H}N_i-y_{NS}\overline{N_i}S\sigma+{\rm H.c.}-V(H,\sigma),\label{eq:lag}
\end{eqnarray}
where $i\in[1,2],\alpha\in[e,\mu,\tau]$ and the most general scalar potential is given as
\begin{eqnarray}
    V(H,\sigma)=-\mu_h^2H^\dagger H+\lambda_h(H^\dagger H)^2+\frac{1}{2}\mu_\sigma^2\sigma^2+\frac{1}{4}\lambda_\sigma\sigma^4+\lambda_{h\sigma}H^\dagger H\sigma^2,\label{eq:pot}
\end{eqnarray}
where we choose $\mu_\sigma^2>0$ such that $\sigma$ does not obtain any VEV. The Higgs field can be parameterized after electroweak symmetry breaking as $H=\left(0~~~(h+v_h)/\sqrt{2}\right)$ with $v_h$ being the VEV of the SM Higgs.

Depending on the reheating temperature of the Universe, the DM relic can be realized in two distinct scenarios: $i$) for reheating temperature, $T_{\rm rh}>m_{N_1}$, and $ii$) for reheating temperature, $T_{\rm EW}<T_{\rm rh}\ll m_{N_1}$, where $m_{N_1}$ is the mass of lightest RHN $N_1$. 

\underline{\textit{Case-i) $T_{\rm rh}>m_{N_1}$}:} As the $T_{\rm rh}$ is larger than the mass of $N_1$, it can be produced abundantly from the thermal bath through the $LHN$ coupling.  As such, the out-of-equilibrium decay of $N_1$ in the early Universe can also give rise to a net lepton asymmetry. Moreover, if the coupling between $\sigma$ and SM Higgs, $\lambda_{h\sigma}$ is large enough  (typically $\lambda_{h\sigma}>1$), it will trigger a first-order electroweak phase transition (FOEWPT). This will result in stochastic gravitational waves, potentially observable in future GW experiments. Moreover, the freeze-out relic of $\sigma$ is small. Therefore, any late decay of $\sigma$ will give a negligible contribution to the DM relic. On the other hand, if $\lambda_{h\sigma}$ (typically $\lambda_{h\sigma}\ll1$) coupling is small, $\sigma$ freezes out from the thermal bath with a larger abundance. Its late decay can then contribute to the DM relic density as well as additional relativistic degrees of freedom (d.o.f) ($\Delta{N}_{\rm eff}$) in the form of light neutrinos. Such a scenario may yield distinctive signatures testable in future CMB experiments. By demanding that $\Delta{N}_{\rm eff}\lesssim0.4$, we will show that late decay of $\sigma$ cannot give rise to the correct relic abundance of $S$. Thus, irrespective of the values of $\lambda_{h\sigma}$, the contribution of $\sigma$ to the DM relic is negligible. In this scenario, the relic of DM is dominantly produced by $N_1$ decay (\textit{i.e.,} $N_1\rightarrow S\sigma$). Even if $h\rightarrow SS$ is allowed at one loop, its contribution to $S$ abundance is negligible in comparison to $N_1$ decay. This is because the production of $S$ from Higgs decay is proportional to $y_{NS}^4$, whereas the production of $S$ from $N_1$ decay is proportional to $y^2_{NS}$. 

\underline{\textit{Case-ii) $T_{\rm EW}<T_{\rm rh}\ll m_{N_1}$}:} In this scenario, $N_1$ cannot be produced from the thermal plasma. However, the DM can be produced from the SM Higgs decay at one loop level. As discussed in the previous paragraph, this scenario also may yield distinctive signatures testable in future CMB and GW experiments. Similar to the above discussion, the contribution of $\sigma$ decay to the DM relic remains negligible. 

We will study the phenomenologies of these two scenarios: case-$i$ and case-$ii$ in detail in the following sections.

\section{Phenomenology with $T_{\rm rh}>m_{N_1}$}\label{sec:sddmrelic}

As discussed above, when the reheat temperature of the Universe, $T_{\rm rh}>m_{N_1}$, the RHNs can be abundantly produced in the early Universe through the $NLH$ coupling. Through the same coupling, the \textit{CP} violating out-of-equilibrium decay of $N\rightarrow LH$, can generate a net lepton asymmetry, which then gets converted to baryon asymmetry of the Universe\footnote{The simultaneous decay of RHNs to visible sector and dark sector can give rise to asymmetric dark matter scenarios, see \textit{e.g.} \cite{Falkowski:2011xh,Borah:2024wos,Mahapatra:2023dbr}. In the present scenario, $S$ being a Majorana fermion, asymmetry in $S$ does not exist.}. The \textit{CP} asymmetry can be defined as
\begin{eqnarray}
\epsilon=\frac{\Gamma(N\rightarrow LH)-\Gamma(N\rightarrow \bar{L}H^\dagger)}{\Gamma(N\rightarrow LH)+\Gamma(N\rightarrow \bar{L}H^\dagger)}.
\end{eqnarray}
A very high-scale seesaw places the viability of the leptogenesis scenario beyond the reach of current experimental facilities. However, the leptogenesis scale can be lowered by considering resonant enhancement of the \textit{CP} asymmetry \cite{Pilaftsis:2003gt}. In standard resonant leptogenesis, a mild mass degeneracy between the RHNs $N_1, N_2$ results in a resonant enhancement of the CP asymmetry via contributions from the self-energy diagrams. The \textit{CP} violating out-of-equilibrium decay of $N_1$, and $N_2$ to $L_\alpha, H$ generates the lepton asymmetry in the early Universe in different flavors. The key parameters in the leptogenesis are \{ $m_{N_1}$, $\delta M$, $\gamma$\}, where $\delta{M}=m_{N_2}-m_{N_1}$ and $\gamma$ is the complex rotation angle of the orthogonal matrix used in the Casas-Ibarra (CI) parametrization in Eq \ref{eq:cipara}. In the flavored regime, the evolution of the lepton asymmetry is governed by the density matrix formalism, with the diagonal elements representing flavor-specific asymmetries and the off‑diagonal elements encoding quantum coherence between flavors. For TeV‑scale leptogenesis, rapid thermal interactions erase this coherence, strongly damping the off‑diagonal terms. Consequently, the density matrix equations simplify to a set of flavor‑diagonal Boltzmann equations, one for each lepton flavor asymmetry. This simplification remains valid when $m_{N_i}\ll10^9$ GeV \cite{Blanchet:2011xq,Moffat:2018wke}. In our analysis, we have incorporated all relevant lepton-number-violating processes, including both Yukawa-mediated and gauge-mediated channels. The gauge processes, in particular, are crucial, as their typically strong interaction rates keep them in equilibrium over a wide temperature range, leading to a significant enhancement of washout effects. The produced lepton asymmetry is subsequently converted into baryon asymmetry via electroweak sphalerons. This process must be completed before $T_{\rm sph} >131.7~\mathrm{GeV}$~\cite{DOnofrio:2014rug}.
 The \textit{CP} asymmetry parameter for the decay of $N_i$ into $L_\alpha$ is given as \cite{DeSimone:2007edo,Huang:2024azp}
\begin{eqnarray}
\epsilon^i_{\alpha\alpha}&=&-\sum_{j\neq i} \frac{ {\rm Im}\left[ (y^\dagger)_{i\alpha}(y)_{\alpha j} (y^\dagger y)_{ij} +\frac{m_{N_i}}{m_{N_j}} (y^\dagger)_{i\alpha} y_{\alpha j} (y^\dagger y)_{ji} \right] }{(y^\dagger y)_{ii} (y^\dagger y)_{jj}}\times \left(\frac{(m_{N_i}^2-m_{N_j}^2)m_{N_i}\Gamma_j}{(m_{N_i}^2-m_{N_j}^2)^2+m_{N_i}^2\Gamma_j^2} + \right.\nonumber\\&& \left. \frac{(m_{N_i}^2-m_{N_j}^2)m_{N_i}\Gamma_j}{(m_{N_i}^2-m_{N_j}^2)^2+(m_{N_i}\Gamma_i+m_{N_j}\Gamma_j)^2 \frac{{\rm det}[{\rm Re} (y^\dagger y)]}{(y^\dagger y)_{ii} (y^\dagger y)_{jj}} }\right).
\end{eqnarray}
Assuming two degenerate RHNs, the $3\times2$ Yukawa coupling matrix can be parameterized as
\begin{eqnarray}
    y=\frac{\sqrt{2}}{v_h}U_{\rm PMNS}\sqrt{\hat{m}_\nu}.\mathcal{R}^T.\sqrt{\hat{M}},\label{eq:cipara}
\end{eqnarray}
where $U_{\rm PMNS}$ is the Pontecorvo–Maki–Nakagawa–Sakata matrix, $\hat{m}_\nu={\rm diag}(0,m_2,m_3)$ is the diagonal light neutrino mass matrix, $\hat{M}={\rm diag}(m_{N_1}, m_{N_2})$ is the diagonal heavy RHN matrix and $\mathcal{R}$ is the orthogonal rotational matrix defined as
\begin{eqnarray}
    \mathcal{R}=\begin{pmatrix}
        0 & \cos\gamma& \sin\gamma\\
        0 & -\sin\gamma & \cos\gamma
    \end{pmatrix},
\end{eqnarray}
where $\gamma$ is the complex rotation angle.
The baryon asymmetry then can be written in terms of the lepton asymmetries as
\begin{eqnarray}
    \eta_B=7.04\frac{C}{f}\left( Y_{\Delta e}+Y_{\Delta\mu}+Y_{\Delta\tau} \right)=7.04\frac{C}{f}Y_{\Delta} ,
\end{eqnarray}
where $C=28/79$ is the fraction of lepton asymmetry converted into a baryon asymmetry by sphaleron processes, $f=g^*_s/g^*_0=28$\footnote{$g^*_s=106.75+\frac{7}{8}\times2+1=109.5$ is the relativistic degrees of freedom (d.o.f) at the onset of leptogenesis in our scenario, $g^*_0$ = 3.91 is the present day relativistic d.o.f.} is the dilution factor calculated assuming standard photon production from the onset of leptogenesis till recombination, and $Y_\Delta$ denotes the total lepton asymmetry. Here $Y_{\Delta e},Y_{\Delta \mu},Y_{\Delta \tau}$ can be obtained by solving the relevant Boltzmann equations as discussed in \ref{app:lepto}.

Apart from $N\rightarrow LH$, $N$ can also decay to $S\sigma$ and produce a freeze-in population of DM $S$. Additionally, the decay of $\sigma$ to $S\nu$ via $N-\nu$ mixing can also produce the DM relic via the SuperWIMP mechanism. Therefore, the relic of $S$ can be affected by the $\lambda_{h\sigma}$ coupling, thereby giving two distinct cases of BSM phenomenologies to arise: A) large $\lambda_{h\sigma}$ (typically $\lambda_{h\sigma}>1$) leading to an under-abundant freeze-out population of $\sigma$. In this case $S$ population arise through $N\rightarrow S\sigma$ channel only. B) small $\lambda_{h\sigma}$ (typically $\lambda_{h\sigma}\ll1$). In this case, $\sigma$ can freeze out with a large abundance and subsequently decay to $S$, which can give rise to DM relic along with the decay of $N\rightarrow \sigma S$. The details will be discussed below.

\subsection{Consequences of large $\lambda_{h\sigma}$}\label{sec:largelambda}
\subsubsection{First-order electroweak phase transition (FOEWPT) and stochastic gravitational waves}\label{sec:foptlargelambda}
\begin{figure}[ht]
\centering
\includegraphics[scale=0.4]{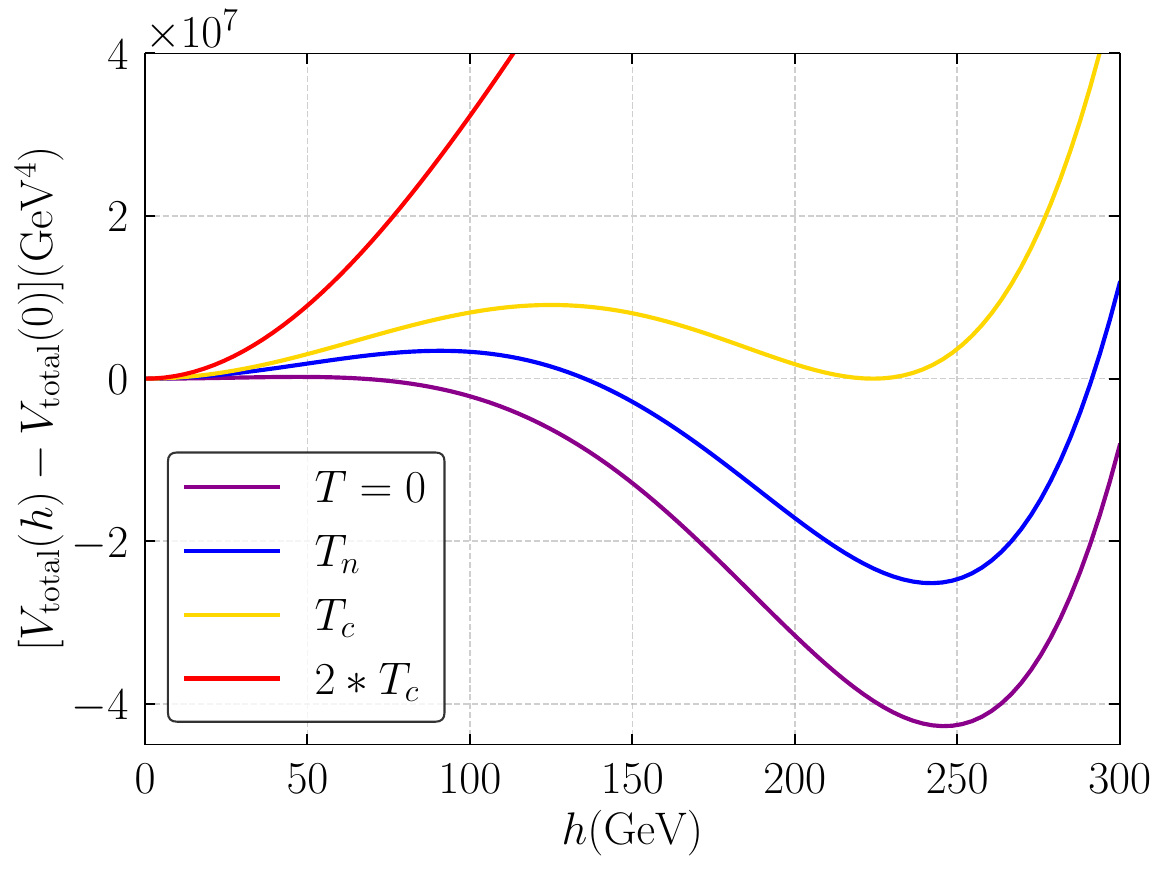}
\includegraphics[scale=0.4]{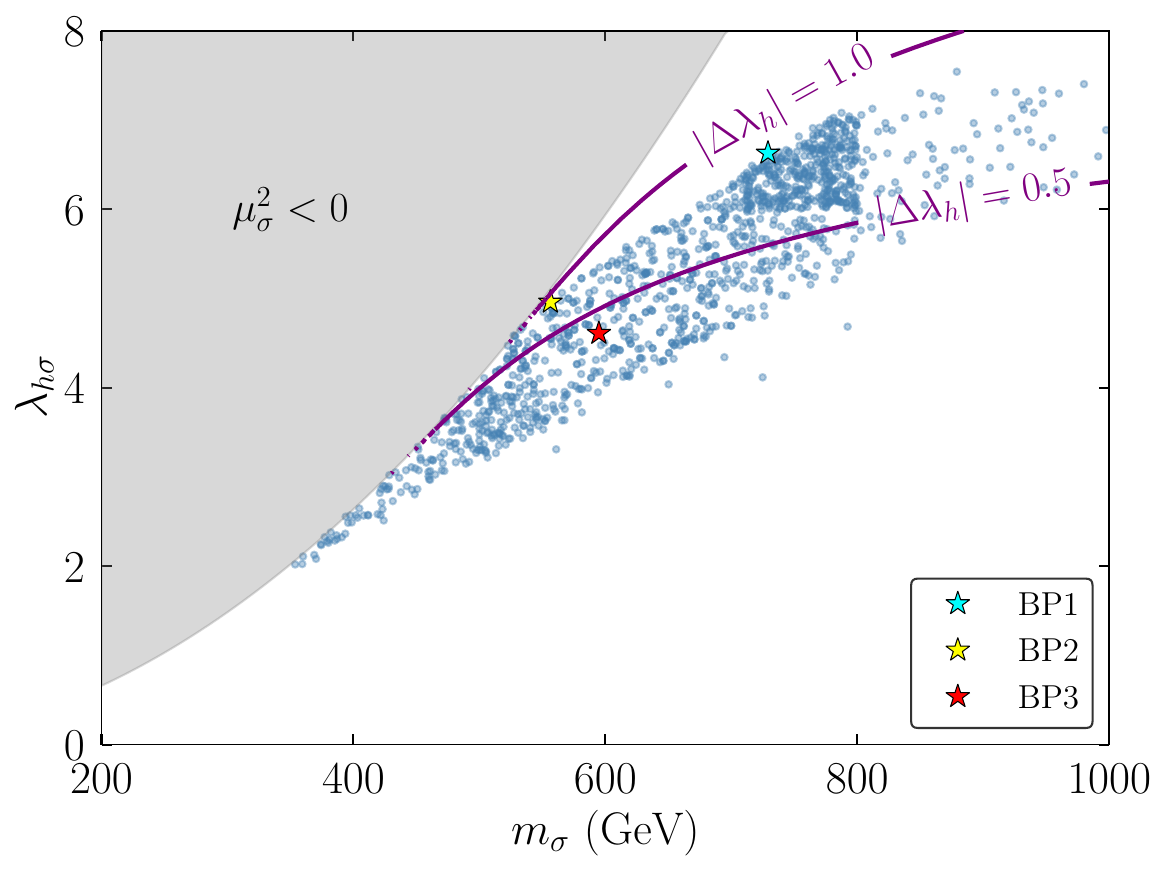}
\caption{[\textit{Left}:] Effective potential profile for BP1 from Table \ref{tab:tab1} at different temperatures. [\textit{Right}:] Parameter space (shown by light blue dotted points) of first-order EWPT in the plane of $\lambda_{h\sigma}$ vs $m_\sigma$. The purple contours 
show $|\Delta\lambda_h(h=0)|$ [Eq.~(\ref{eq:deltalambda})], the one-loop 
singlet correction to the Higgs quartic coupling, following Ref.~\cite{Curtin:2014jma}. The gray shaded region corresponds to 
$\mu_\sigma^2 < 0$, which is excluded in our scenario. The colored stars 
denote the benchmark points of Table~\ref{tab:tab1}.}
\label{fig:foptpar}
\end{figure}

As discussed above $\sigma$ does not obtain any VEV and only SM Higgs obtains a VEV: $v_h$. The $\sigma$ mass gets a contribution from SM Higgs and is given as $m_\sigma=\sqrt{\mu_\sigma^2+\lambda_{h\sigma}v_h^2}$. The SM Higgs mass is $m_h=\sqrt{2}\mu_h=125$ GeV and $\lambda_h=0.129$. The only free parameters in the scalar sector are \{$m_\sigma,\lambda_\sigma,\lambda_{h\sigma}$\}. The EWPT can be made first-order with the help of this singlet scalar $\sigma$ through the coupling $\lambda_{h\sigma}$ \cite{Curtin:2014jma,Dey:2025pcs}. Singlet scalar quartic coupling, $\lambda_\sigma$ does not play any role in the mass of $\sigma$, it only contribute to its thermal mass. Following the approach of Ref.~\cite{Curtin:2014jma}, we set $\lambda_\sigma = 0$ to maximize the FOPT parameter space,  as nonzero $\lambda_\sigma$ only contributes to the thermal mass of $\sigma$ and weakens the phase transition. The bounded-from-below conditions $\lambda_h > 0$, $\lambda_\sigma \geq 0$, and $\lambda_{h\sigma} > -2\sqrt{\lambda_h \lambda_\sigma}$~\cite{Kannike:2016fmd,Curtin:2014jma} are satisfied throughout our parameter space. Now, the only free parameters determining the FOPT are \{$m_\sigma,\lambda_{h\sigma}$\}. The finite temperature effective scalar potential is given in \ref{app:fopt}. We implement the model in {\tt CosmoTransitions} \cite{Wainwright:2011kj} and obtain the parameter space of the first-order EWPT in the $\lambda_{h\sigma}-m_\sigma$ plane, as shown in the \textit{right} panel of Fig. \ref{fig:foptpar} with light blue color.

To assess the perturbative reliability of our analysis, we follow  Ref.~\cite{Curtin:2014jma} and examine the one-loop singlet contribution  to the Higgs quartic coupling,
\begin{equation}
\Delta\lambda_h(h) = \frac{\lambda_{h\sigma}^2}{16\pi^2}\left[\log\left(1 + 
\frac{(h^2 - v_h^2)\lambda_{h\sigma}}{m_\sigma^2}\right) - \frac{3}{2}\right].
\label{eq:deltalambda}
\end{equation}
Two-loop corrections scale as $\sim (\Delta\lambda_h)^2$, so the perturbative 
expansion is reliable when $|\Delta\lambda_h|$ is not too large. In the \textit{right} panel of
Fig.~\ref{fig:foptpar}, we show contours of $|\Delta\lambda_h(h=0)|$ 
in the $\lambda_{h\sigma}$--$m_\sigma$ plane, overlaid on the FOEWPT parameter space. We show three benchmark points (BPs), which will be used in later discussions, with colored stars as listed in Table \ref{tab:tab1}. BP3, with $\lambda_{h\sigma} = 4.61$, lies below the $|\Delta\lambda| = 0.5$ contour, where the one-loop analysis is well under control. BP2 ($\lambda_{h\sigma} = 4.96$) lies near the $|\Delta\lambda| \simeq 0.5$ boundary, while BP1 ($\lambda_{h\sigma} = 6.63$) sits in the region $|\Delta\lambda| \sim 0.8$, where higher-order corrections may become relevant. A dedicated two-loop analysis would refine the quantitative predictions for such large couplings, which is beyond the scope of the present work, but the qualitative conclusion remains robust.
\begin{figure}[ht]
    \centering
    \includegraphics[scale=0.42]{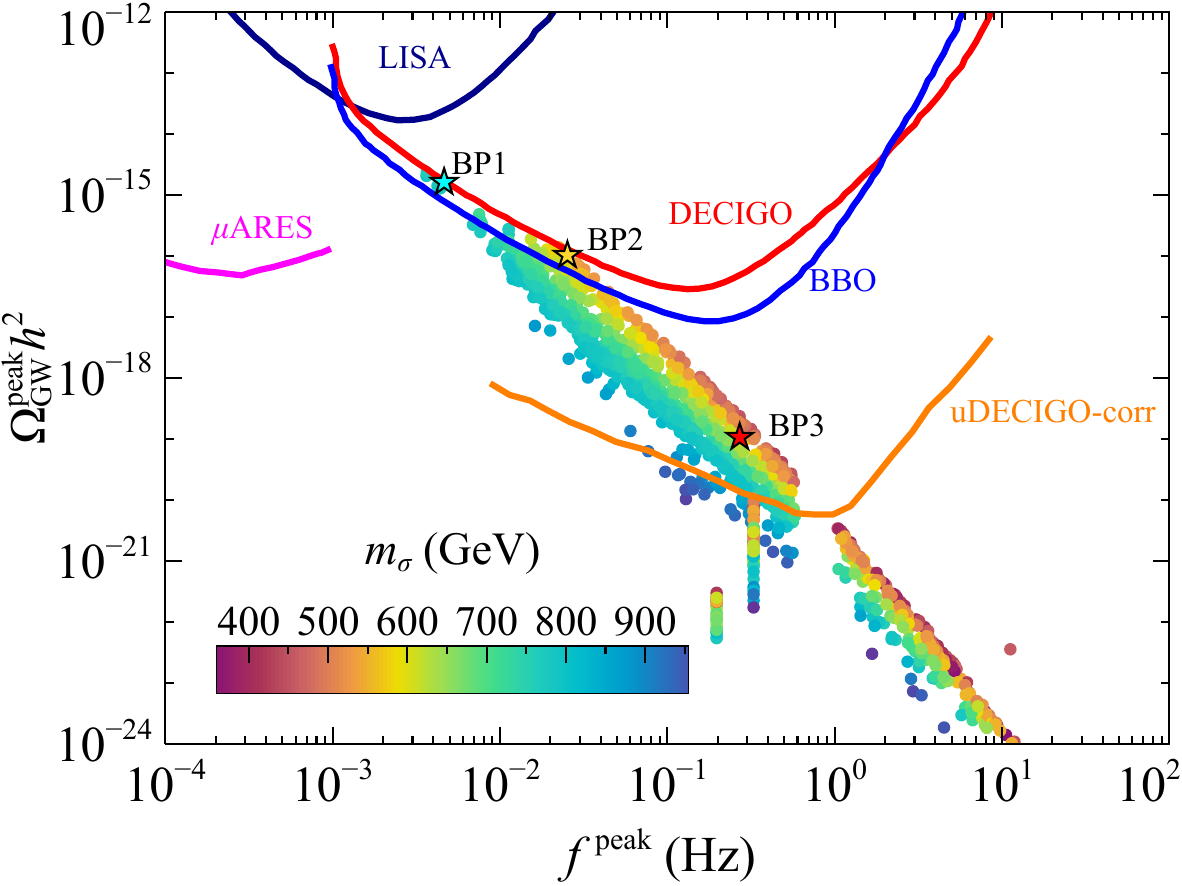}
    \caption{Peak GW amplitude as a function of peak frequency.}
    \label{fig:gwamp}
\end{figure}
In the gray shaded region $\mu_\sigma^2<0$, the singlet scalar would obtain a non-zero VEV and destabilize the DM due to $S-N$ mixing. Therefore, our region of interest is $\mu_\sigma^2>0$ and the black points represent the FOEWPT. This restricts the mass and coupling to be in the range $m_\sigma\in[350,1000] {\rm ~GeV}, \lambda_{h\sigma}\in[2, 7.5]$. With this large $\lambda_{h\sigma}$, $\sigma$ will always be in equilibrium and freeze out with negligible relic abundance and will not be a DM candidate, as will be discussed in the next section.

In the \textit{left} panel of Fig.~\ref{fig:foptpar}, we illustrate the temperature dependence of the effective potential for the BP1 from Table \ref{tab:tab1}. At the critical temperature $T_c$, a barrier develops between two degenerate minima, shown by the yellow curve. The shape of the potential at the nucleation temperature $T_n$ is depicted by the blue curve. At a much higher temperature, $T = 2\,T_c$, the field remains in the symmetric phase, as indicated by the red curve. Finally, at zero temperature, the field settles in the global minimum, shown by the dark magenta curve.

We then compute the peak amplitude and peak frequencies for these FOPT points as shown in Fig. \ref{fig:gwamp}. The color code denotes the mass of $\sigma$. Sensitivity reaches of different experiments like DECIGO \cite{Kawamura:2020pcg}, BBO \cite{Yagi:2011wg}, uDECIGO-corr \cite{Kawamura:2020pcg}, LISA \cite{LISA:2017pwj}, and $\mu$ARES \cite{Sesana:2019vho} are shown with different color contours. The peak frequency spans from mHz to a few Hz. The amplitude remains in the sensitivities of DECIGO, BBO, and uDECIGO-corr.

\begin{table}[h]
\centering
\begin{tabular}{|c|c|c|c|c|c|c|} 
 \hline
 BPs & $m_\sigma$ (GeV) & $\lambda_{h\sigma}$&$T_c$ (GeV) &$T_n$ (GeV)& $\alpha$ &$\beta/\mathcal{H}$\\ [0.5ex] 
 \hline
  BP1 & 729.1 &6.63 &139.3 &94.45 &0.024 &255.79 \\[0.7ex] 
 \hline
 BP2 & 556.34 &4.96 &128.02 &108.51 &0.024 &1219.19\\[0.7ex]  
 \hline
 BP3 &594.86  &4.61 &147.39 &143.86 &0.0093 &9838.08\\[0.7ex]  
 \hline
\end{tabular}
\caption{Details of the parameters for the three BPs shown in Fig. \ref{fig:foptpar} and \ref{fig:gwamp}. Here $T_c$ denotes the critical temperature, $T_n$ denotes the nucleation temperature, $\alpha$ is the strength of FOPT defined as the ratio of vacuum energy density to radiation energy density, and $\beta$ represents the inverse duration of phase transition.}\label{tab:tab1}
\end{table}

\subsubsection{Freeze-in production of dark matter from $N_1$ decay}\label{sec:DMleptoLagelambda}

The singlet fermion $S$, which is even under the lepton parity, acts as a natural candidate for DM in our scenario. It is known that if a singlet fermion is thermalized, then its freeze-out relic remains overabundant in most of the parameter space. Here, we assume that $S$ never thermalizes and its relic is realized via the freeze-in mechanism \cite{McDonald:2001vt,McDonald:2008ua,Hall:2009bx,Konar:2025iuk}. In this case, the freeze-in abundance of $S$ can be obtained from the decay of $N_1,\sigma,H$. The DM relic can be obtained from the $\sigma$ decay after $\sigma$ is frozen out. However, for the choice of $\lambda_{h\sigma}$ required for FOEWPT, $\sigma$ will always freeze out with a small relic abundance and hence its contribution to DM is negligible. Another possibility is annihilation of $\sigma\sigma$ to $SS$ with $N_1$ in the \textit{t-channel} and \textit{u-channel} diagrams. However, the rates of these processes are suppressed since the corresponding rate is proportional to $y_{NS}^4$. We now discuss the possibility of DM production from $N$ decay. As the reheat temperature, $T_{\rm rh}$, is larger than $m_{N_1}$, DM can be produced directly from the decay of $N_1$. If $N_2$ is thermally produced, then its decay also contributes to the relic of $S$. Here, for simplicity, we only consider the decay of $N_1$ to DM, assuming that the $\overline{N_2} S \sigma$ coupling is negligible. Similarly, the SM Higgs decay to $SS$ at one loop also contributes to the DM relic abundance. However, we note that its contribution is negligible compared to $N_1$ decay as its rate is proportional to $y_{NS}^4$. Nevertheless, here we write the full Boltzmann equations (BE) for the evolution of $N_1,\sigma$, and $S$ as

\begin{eqnarray}
    \frac{dY_{N_1}}{dz}=-(D_1+S^t_1+S^s_1)(Y_{N_1}-Y^{\rm eq}_{N_1})-\frac{\langle\Gamma_{N_1\rightarrow \sigma S}\rangle}{z\mathcal{H}(T)} \left(Y_{N_1}-Y_S\frac{Y^{\rm eq}_{N_1}}{Y^{\rm eq}_S}\right) ,\label{eq:N1}
\end{eqnarray}
\begin{eqnarray}
\frac{dY_S}{dz}=\frac{s(T)}{z\mathcal{H}(T)}\Bigg[\langle \sigma v\rangle_{{\sigma\sigma}\rightarrow SS}\left(Y^{2}_{\sigma}-Y_S^2\frac{(Y^{\rm eq}_\sigma)^2}{(Y^{\rm eq}_S)^2}\right) +\frac{\langle\Gamma_\sigma\rangle}{s(T)} \left(Y_{\sigma}-Y_S\frac{Y^{\rm eq}_\sigma}{Y^{\rm eq}_S}\right) +2\frac{\langle\Gamma_h\rangle}{s(T)} Y_h^{\rm eq} + \frac{\langle\Gamma_{N_1\rightarrow \sigma S}\rangle}{s(T)} \left(Y_{N_1}-Y_S\frac{Y^{\rm eq}_{N_1}}{Y^{\rm eq}_S}\right)  \Bigg],\label{eq:BES}
\end{eqnarray}
\begin{eqnarray}
\frac{dY_\sigma}{dz}&=&\frac{s(T)}{z\mathcal{H}(T)}\Bigg[-\langle \sigma v\rangle_{\sigma\sigma\rightarrow {\rm SMSM}}\left(Y_\sigma^2-(Y^{\rm eq}_{\sigma})^2\right)- \langle \sigma v\rangle_{{\sigma\sigma}\rightarrow SS}\left(Y^{2}_{\sigma}-Y_S^2\frac{(Y^{\rm eq}_\sigma)^2}{(Y^{\rm eq}_S)^2}\right) -\frac{\langle\Gamma_\sigma\rangle}{s(T)} \left(Y_{\sigma}-Y_S\frac{Y^{\rm eq}_\sigma}{Y^{\rm eq}_S}\right)+\nonumber\\&&\frac{\langle\Gamma_{N_1\rightarrow \sigma S}\rangle}{s(T)} \left(Y_{N_1}-Y_\sigma\frac{Y^{\rm eq}_{N_1}}{Y^{\rm eq}_\sigma}\right)
\Bigg],\label{eq:BEsigma}
\end{eqnarray}

where $z=m_\sigma/T$, $s(T)=\frac{2\pi^2}{45}g_{*s}T^3$ is the entropy density, $\mathcal{H}(T)=1.66\sqrt{g_*}T^2/M_{\rm pl}$ is the Hubble parameter, $M_{\rm pl}=1.22\times10^{19}$ GeV is the Planck mass.
$\langle\Gamma_{N_1\rightarrow \sigma S}\rangle$ is the thermally averaged decay width for $N_1\rightarrow \sigma S$, $\langle\Gamma_{\sigma}\rangle$ is the thermally averaged decay width for $\sigma\rightarrow S\nu$, $\langle\Gamma_{h}\rangle$ is the thermally averaged decay width for $h\rightarrow SS$, $\langle\sigma v\rangle_{\sigma\sigma\rightarrow SS}$, $\langle\sigma v\rangle_{\sigma\sigma\rightarrow {\rm SMSM}}$ are the thermally averaged cross-sections involving $\sigma$ annihilation. $D_{1}=\langle\Gamma_{N_1\rightarrow LH}\rangle/\mathcal{H}z$, where $\langle\Gamma_{N_1\rightarrow LH}\rangle$ is the thermally averaged decay width for $N_1\rightarrow LH$.  $S^{s,t}_i=\Gamma^{s,t}_i/\mathcal{H}z$ are the scattering terms which include both Yukawa and gauge boson contributions involving the visible sector. The expressions for the decay widths and cross sections are given in \ref{app:decayrate} and \ref{app:crossec}.

The $N_1LH$ coupling is related to the neutrino mass and has to be large for larger masses of $N_1$. Thus, $N_1$ will attain equilibrium from the inverse decay $LH\rightarrow N_1$. This Yukawa coupling can be computed from the neutrino oscillation data using the Casas-Ibarra (CI) parameterization as given in Eq. \ref{eq:cipara}. We remind that the DM ($S$) should not be thermalized. Since it has only one annihilation mode, $SS\rightarrow \sigma\sigma$ (forbidden channel), DM will freeze out with a larger abundance. This is because the rate of this process is suppressed by the mass of $m_{N_1}$. To avoid thermalization, $y_{NS}$ is bounded from above. Assuming that the decay happens around the mass scale of $N_1$ we get
\begin{eqnarray}
    y_{NS}\lesssim 4.3\times10^{-6}\left( \frac{g^d_{*}}{100} \right)^{1/4} \left( \frac{m_{N_1}}{10^6~\rm GeV} \right)^{1/2},
\end{eqnarray}
where $g^d_*$ is the relativistic degrees of freedom at the time of $N_1$ decay.  

The observed relic of DM can be parameterized in terms of comoving number density $Y_S$ and its mass $m_S$ as 
\begin{eqnarray}
\Omega_{\rm DM}h^2\simeq0.118\left(\frac{Y_S}{4.2\times10^{-10}}\right)\left(\frac{m_S}{1~\rm GeV}\right).
\end{eqnarray}

\begin{figure}[ht]
    \centering
    \includegraphics[scale=0.4]{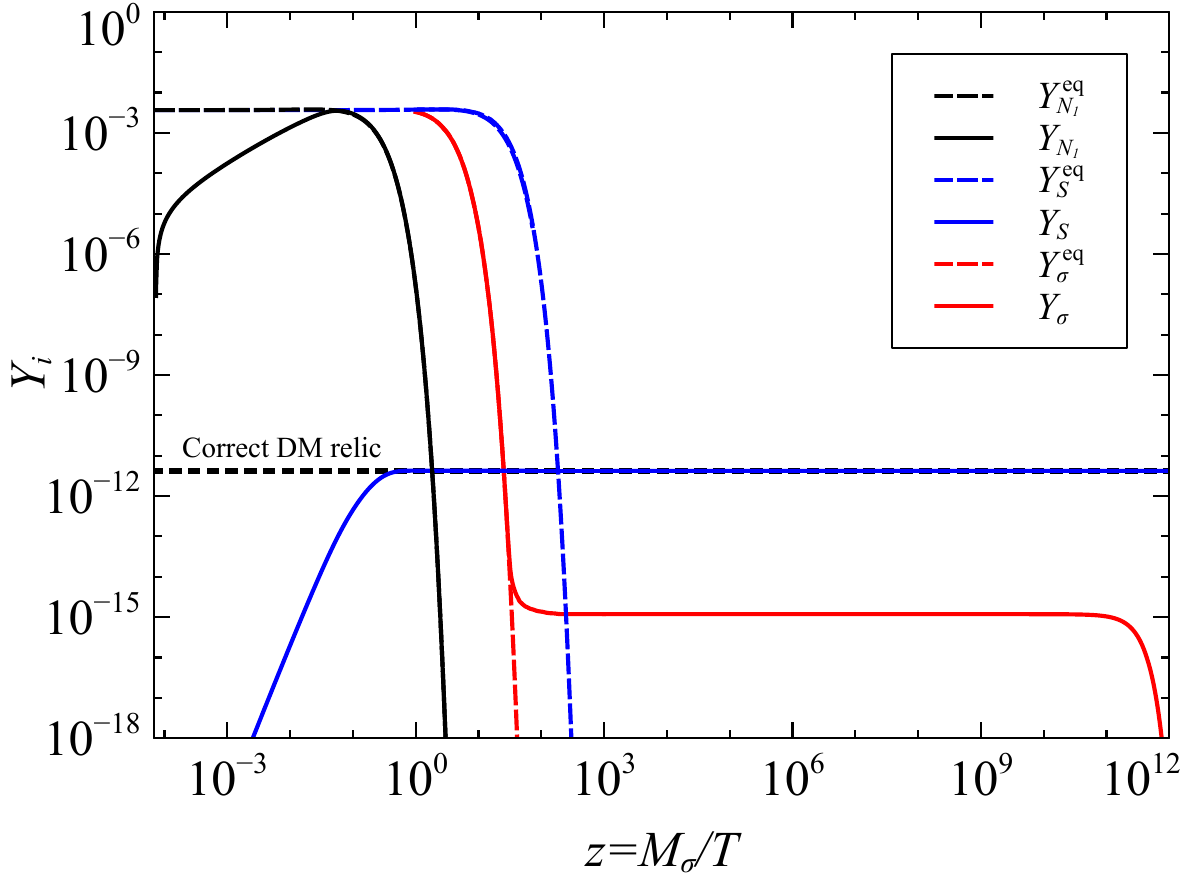}
    \includegraphics[scale=0.4]{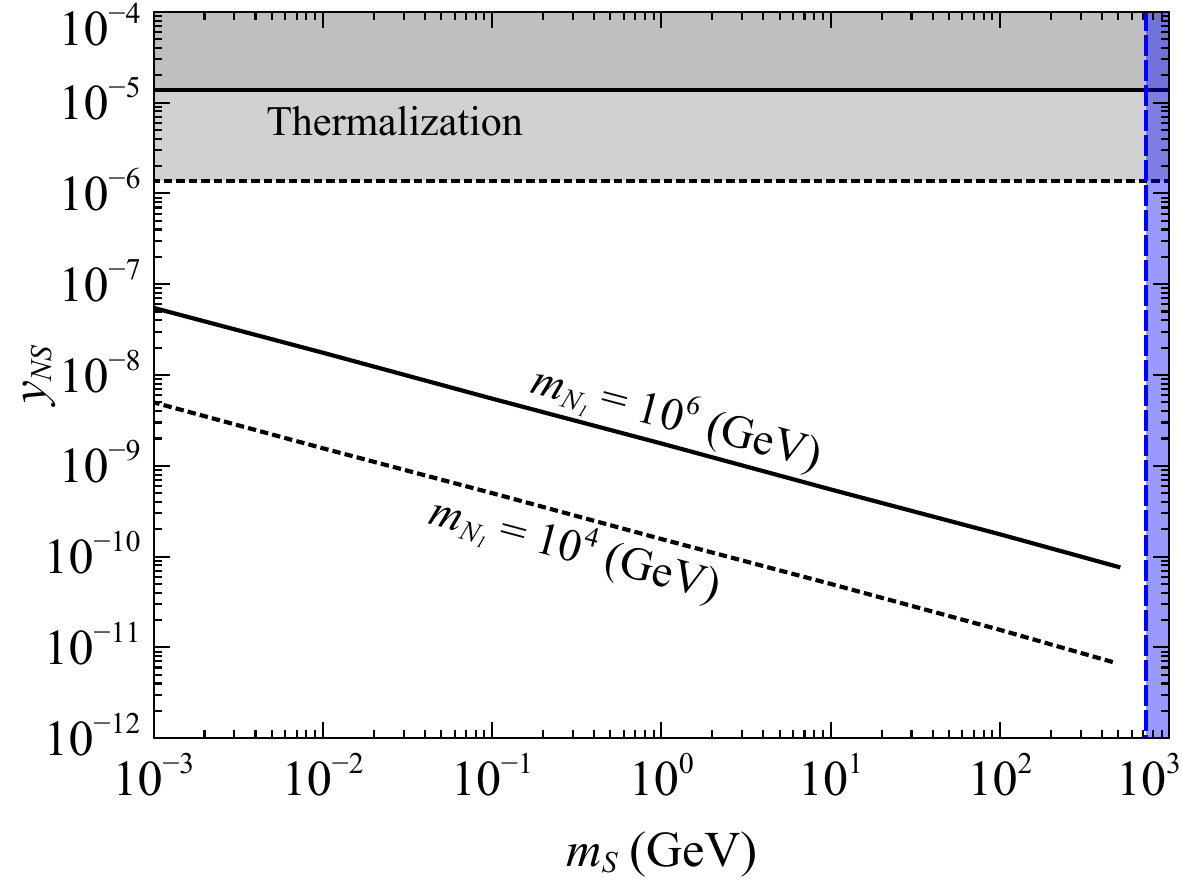}
    \caption{[\textit{Left}:] Cosmological evolution of the abundances of $N_1,\sigma$, and $S$. The corresponding equilibrium abundances are shown with the dashed line. The parameters are fixed at \{$m_{N_1}=10^{4}$ GeV, $m_\sigma=729.1$ GeV, $\lambda_{h\sigma}=6.63$, $m_S=100$ GeV, $y_{NS}=1.56\times10^{-11}$, $\delta{M}=2.1\times10^{-4}$ GeV,$\gamma=0.15-i0.15$\}. [\textit{Right}:] Correct DM relic contours in the plane of $y_{NS}$ vs $m_S$ for fixed $m_{N_1}=10^{6}$ GeV (black solid) and $m_{N_1}=10^{4}$ GeV (black dashed). The correct lepton asymmetry is generated for $\delta{M}=0.521$ GeV for $m_{N_1}=10^6$ GeV and $\delta{M}=2.1\times10^{-4}$ GeV for $m_{N_1}=10^4$ GeV. The rotation angle is fixed at $\gamma=0.15-i0.15$. The region of thermalization of DM is shown with gray shading for these two masses, respectively. The blue shaded region corresponds to $m_S>m_\sigma$.}
    \label{fig:bp1N}
\end{figure}

\begin{figure}[ht]
    \centering
    \includegraphics[scale=0.42]{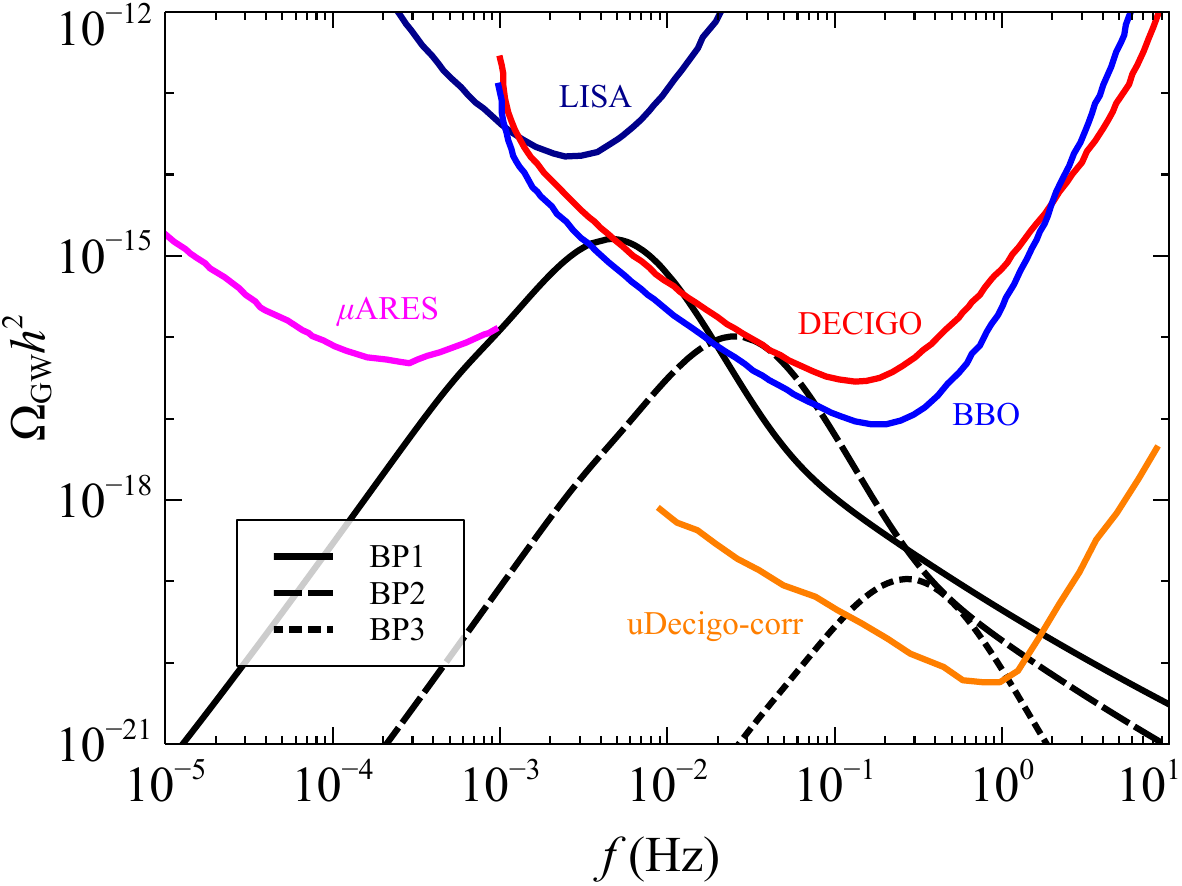}
    \caption{Gravitational wave spectrum from first order electroweak phase transition for BP1, BP2, and BP3 as mentioned in Fig. \ref{fig:gwamp}.}
    \label{fig:gwampspectrum}
\end{figure}

We first solve the BEs \ref{eq:N1}, \ref{eq:BES}, \ref{eq:BEsigma} simultaneously, fixing $m_{N_1}=10^4$ GeV, $\delta{M}=2.1\times10^{-4}$ GeV, $\gamma=0.15-i0.15$, $m_\sigma=729.1$ GeV, $\lambda_{h\sigma}=6.63$, $m_S=100$ GeV, and $y_{NS}=1.56\times10^{-11}$. For this choice of $\lambda_{h\sigma}$, $\sigma$ decouples around $z\sim 37$ and freezes out with a yield of $Y_{\sigma}\sim 1.3\times10^{-15}$. The DM ($S$) gets produced dominantly from the decay of $N_1$ and is shown with the blue solid line in the \textit{left} panel of Fig. \ref{fig:bp1N}. The decay of $\sigma$ to $S$ after it freezes out does not contribute significantly to the abundance of $S$. In this scenario the fraction of $\sigma$ to DM relic is $f_\sigma=\Omega_\sigma/\Omega_{\rm DM}\sim1.97\times10^{-3}$. 

In the \textit{right} panel of Fig. \ref{fig:bp1N}, we show the correct DM relic contours in the plane $y_{NS}$ vs $m_S$ for the two choices of RHN mass as $m_{N_1}=10^4$ and $10^{6}$ GeV. The correct lepton asymmetry can be obtained for these two masses by fixing the mass splitting $\delta{M}=2.1\times10^{-4}$ GeV and $0.521$ GeV respectively, where $\gamma$ is fixed at $0.15-i0.15$. In the upper gray shaded region, the DM will come to thermal equilibrium; hence, it is disallowed in our scenario. The blue shaded region corresponds to $m_S>m_\sigma$.

As previously discussed, large values of $\lambda_{h\sigma}$ will lead to FOEWPT, resulting in stochastic GWs. In Fig. \ref{fig:gwampspectrum}, we show the GW spectrum corresponding to the same benchmark point as in Fig. \ref{fig:bp1N}, which we call BP1. This is shown as a cyan star in Fig. \ref{fig:gwamp}. We also show the spectrum for another two choices of \{$\lambda_{h\sigma},m_\sigma$\}, as shown with yellow and red stars in Fig. \ref{fig:gwamp}. The details of the BPs are given in Table \ref{tab:tab1}. The peak lies within the projected sensitivities of the BBO and DECIGO experiments, while the low and high frequency tails extend into the sensitivity ranges of $\mu$ARES and uDECIGO-corr, respectively.

\subsection{Consequences of small $\lambda_{h\sigma}$}\label{sec:smalllambda}

\subsubsection{Freeze-in and SuperWIMP production of DM}\label{sec:freezeinandsuperwimp}

\begin{figure}[ht]
    \centering
    \includegraphics[scale=0.3]{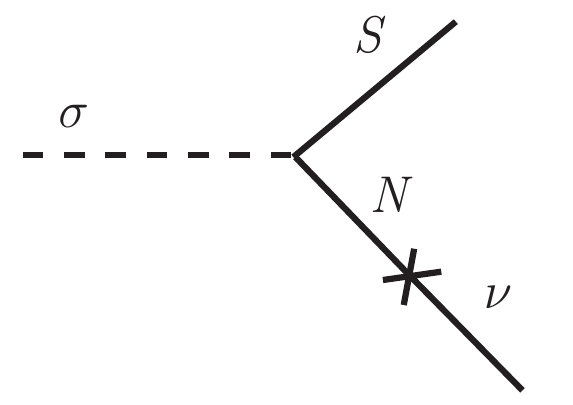}
    \caption{Feynman diagram of $\sigma$ decay to $S$ and $\nu$.}
    \label{fig:sigmadecay}
\end{figure}
If $\lambda_{h\sigma}$ coupling is small compared to the previous case, $\sigma$ freezes out with a relatively larger abundance. In this case, the late decay of $\sigma\rightarrow S\nu$, as shown in Fig. \ref{fig:sigmadecay}, can also produce some amount of DM, which is known as the SuperWIMP contribution, $\Omega_{\rm SuperWIMP}h^2$. The component of DM generated from $N_1$ decay is known as the freeze-in component, $\Omega_{\rm f.i.}h^2$. Thus, the total DM relic abundance is found to be 
\begin{eqnarray}
    \Omega_{\rm DM}h^2=\Omega_{\rm f.i.}h^2+\Omega_{\rm SuperWIMP}h^2.
\end{eqnarray}

 Now the late decay of $\sigma$ not only contributes to DM but also enhances the number of relativistic d.o.f. by injecting additional $\nu_L$ from the same decay mode. This happens if the decay occurs after the SM neutrino decoupling, which we assume to be $T_{\nu}^{\rm dec}\sim1$ MeV and before the CMB epoch: $T_{\rm CMB}\sim0.26$ eV.

\subsubsection{$\Delta N_{\rm eff}$ from late decay of $\sigma$}\label{sec:neff}

If the decay $\sigma\rightarrow S\nu$ happens after the SM neutrinos decouple from the thermal bath, then it can give rise to additional light d.o.f.  In the absence of these light d.o.f., the SM gives a precise prediction for $N_{\rm eff}$, namely $N_{\rm eff}^{\rm SM}$=3.045 \cite{Mangano:2005cc,Grohs:2015tfy,deSalas:2016ztq}. The additional contribution to the effective number of relativistic species at the CMB epoch from $\sigma$ decay is given as
\begin{eqnarray}
    \Delta N_{\rm eff}=N_{\rm eff}-N_{\rm eff}^{\rm SM}=N_{\nu}\frac{\rho_{\nu_L}}{\rho_{\nu_L}^{\rm SM}}\Bigg|_{T_{\rm CMB}},
\end{eqnarray}
where $N_\nu=3$ is the number of generations of SM neutrinos, $\rho_{\nu_L}^{\rm SM}=2\frac{7}{8}\frac{\pi^2}{30}T_{\nu_L}^4$ is the energy density of the SM neutrino, $\rho_{\nu_L}$ is the late time produced neutrino energy density. The DESI 2024 \cite{DESI:2024mwx} data put an upper bound on  $\Delta N_{\rm eff}$ as 0.39. Planck 2018 \cite{Planck:2018vyg} put a constraint as  $\Delta N_{\rm eff}<$ 0.285. The ACT 2025 \cite{ACT:2025tim} data constrain  $\Delta N_{\rm eff}$  to be $\Delta N_{\rm eff}<$0.32. The combined Planck and ACT \cite{ACT:2025tim} data further tighten this constraint to be $\Delta N_{\rm eff}<$0.14. The future CMB experiment SPT-3G \cite{SPT-3G:2019sok} is expected to be sensitive down to $\Delta N_{\rm eff}$= 0.1. The CMB-S4 \cite{Abazajian:2019eic} can probe down to 0.06, whereas CMB-HD \cite{CMB-HD:2022bsz} can probe  $\Delta N_{\rm eff}$ down to 0.027.

The evolution of the energy density of these late-time produced neutrinos is governed by the following BE
\begin{eqnarray}
    \frac{d\rho_{\nu_L}}{dz}=-\frac{4\beta\rho_{\nu_L}}{z}+\frac{1}{z\mathcal{H}(z)}\langle E\Gamma_\sigma\rangle Y_{\sigma}s(z),\label{eq:rhonul}
\end{eqnarray}
where
\begin{eqnarray}
    \beta=1+\frac{T}{3g_s(T)}\frac{dg_{s}}{dT}.
\end{eqnarray}
Here $g_s$ is the effective number of d.o.f. contributing to the entropy density. The thermally averaged energy transfer rate is given as \cite{Biswas:2022vkq,Borah:2025cqj}
\begin{eqnarray}
    \langle E\Gamma_\sigma\rangle=g_S g_\nu \frac{|\mathcal{M}|^2_{\sigma\rightarrow S\nu}}{32\pi}\frac{(m_\sigma^2-m_S^2)^2}{m_\sigma^4},
\end{eqnarray}
where the matrix-squared element for this decay process is given as
\begin{eqnarray}
    |\mathcal{M}|^2_{\sigma\rightarrow S\nu}=\frac{y_{NS}^2y^2 v_h^2}{2m_{N_1}^2}(m_\sigma^2-m_S^2)
\end{eqnarray}
\begin{figure}[ht]
    \centering
    \includegraphics[scale=0.42]{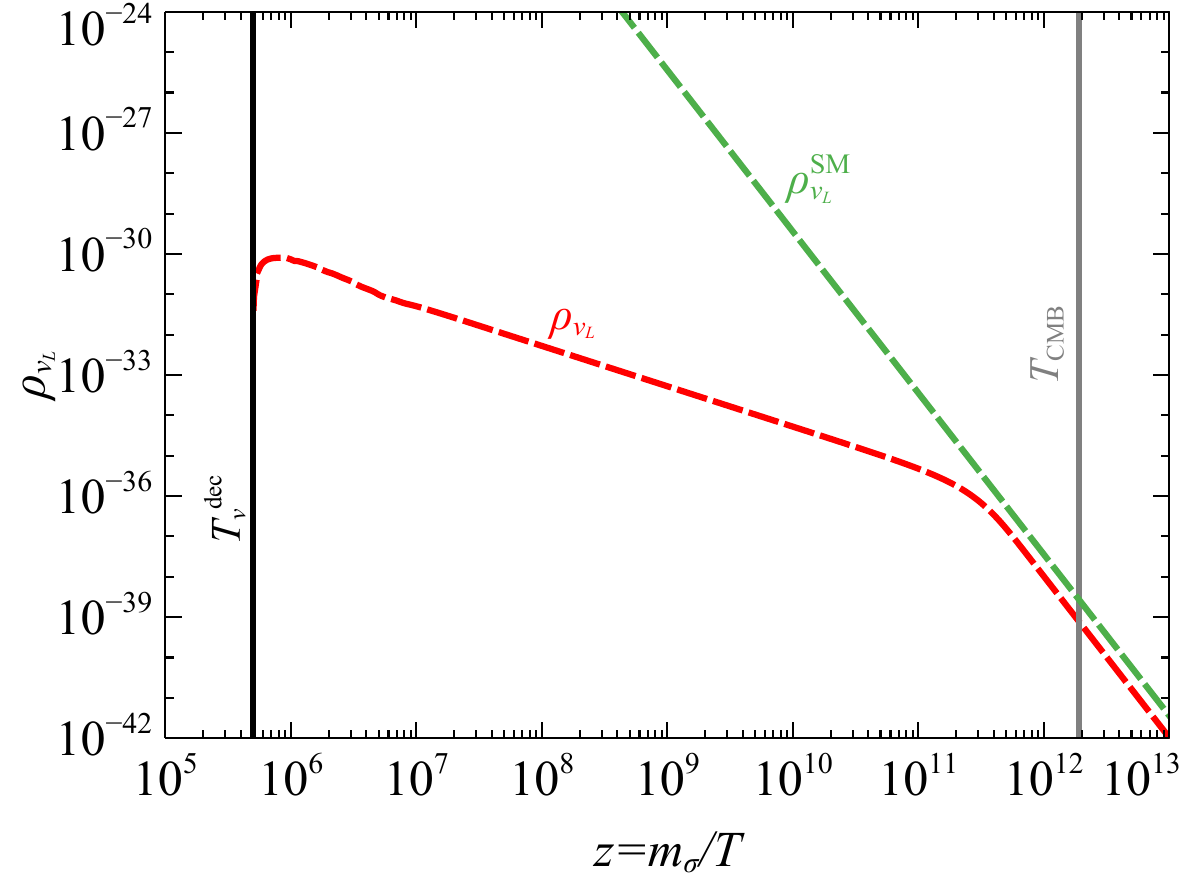}
    \caption{The cosmological evolution of the energy density of the neutrino from $\sigma$ decay (red dashed line) and the SM neutrino (green dashed line) is shown as a function of $z(=m_\sigma/T)$.}
    \label{fig:rhonuL}
\end{figure}
\begin{figure*}[tbh]
    \centering
\includegraphics[scale=0.4]{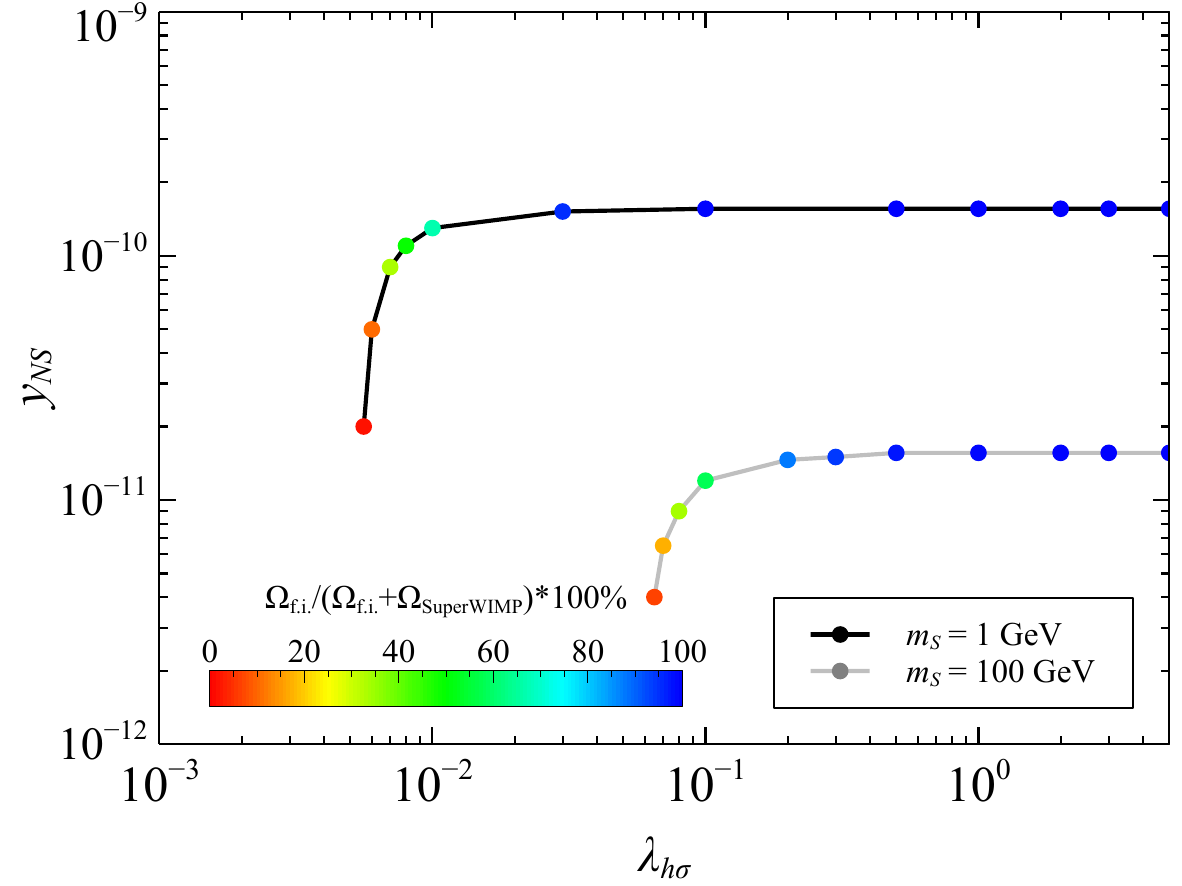}
\includegraphics[scale=0.4]{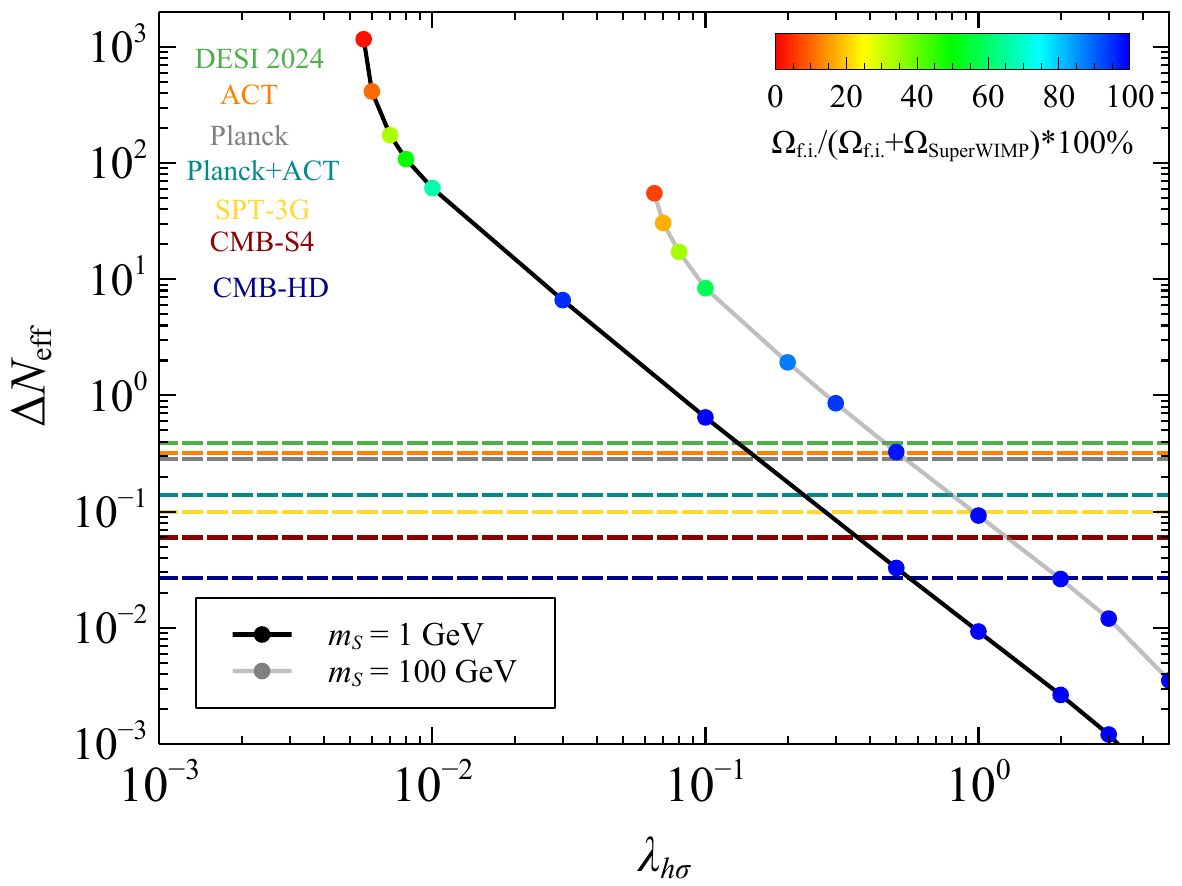}
    \caption{[\textit{Left:}] Correct relic contours in the $y_{NS}-\lambda_{h\sigma}$ plane for two DM masses: $m_{S}$=1 GeV (black solid line) and $m_{S}$=100 GeV (gray solid line). The other parameters are fixed as \{$m_\sigma=500$ GeV, $m_{N_1}=10^4$ GeV, $\delta{M}=2.1\times10^{-4}$ GeV, $\gamma=0.15 - i 0.15$\}. The color code denotes the fraction of freeze-in DM relic from $N_1$ decay to the total DM relic. [\textit{Right:}] The same points in the $\Delta{N}_{\rm eff}-\lambda_{h\sigma}$ plane with same color code.}
    \label{fig:ynsVSlambda1}
\end{figure*}

Now to obtain the energy density of these additional neutrinos, we need to solve Eq. \ref{eq:rhonul}, together with Eqs. \ref{eq:N1}, \ref{eq:BES}, \ref{eq:BEsigma}.
\begin{figure*}[tbh]
    \centering
\includegraphics[scale=0.4]{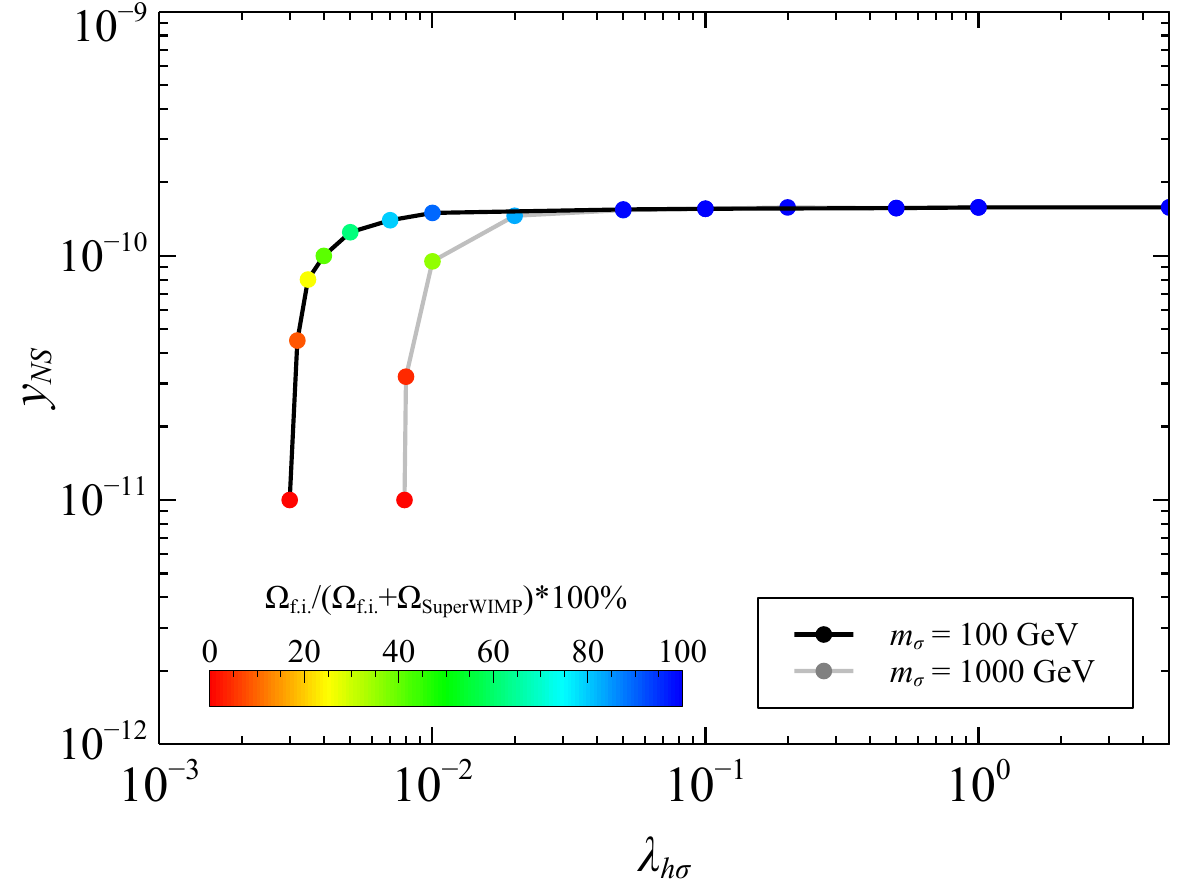}
\includegraphics[scale=0.4]{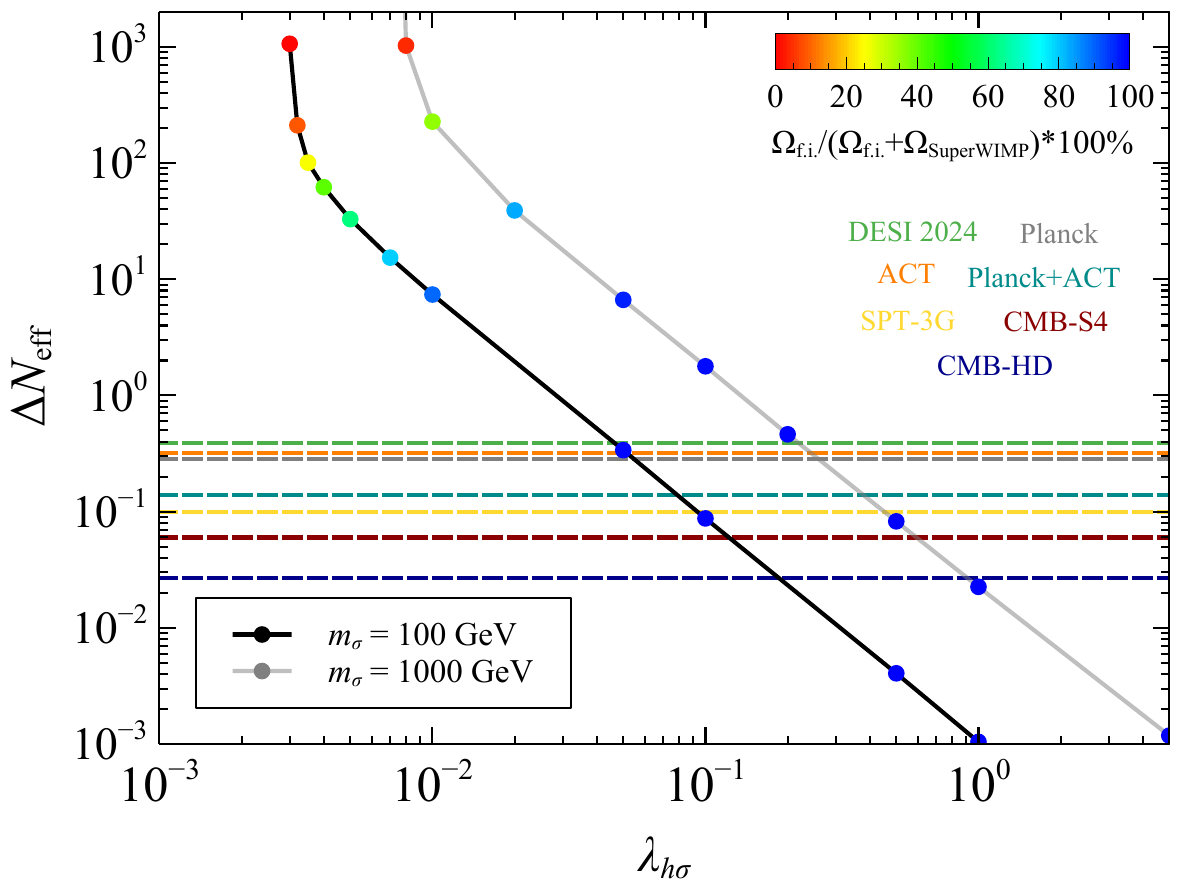}
    \caption{[\textit{Left:}] Correct relic contours in the $y_{NS}-\lambda_{h\sigma}$ plane for two $\sigma$ masses: $m_{\sigma}$=100 GeV (black solid line) and $m_{\sigma}$=1000 GeV (gray solid line). The other parameters are fixed as \{$m_S=1$ GeV, $m_{N_1}=10^4$ GeV, $\delta{M}=2.1\times10^{-4}$ GeV, $\gamma=0.15 - i 0.15$\}. The color code denotes the fraction of freeze-in relic to the total DM relic. [\textit{Right:}] The same points in the $\Delta{N}_{\rm eff}-\lambda_{h\sigma}$ plane. }
    \label{fig:ynsVSlambda2}
\end{figure*}
 We fix the parameters as $m_{N_1}=10^4$ GeV, $\delta{M}=2.1\times10^{-4}$ GeV, $\gamma=0.15-i0.15$, $m_S=100$ GeV, $y_{NS}=1.5\times10^{-11}$, $m_\sigma=500$ GeV, and $\lambda_{h\sigma}=0.3$. In this case, $\sigma$ contribution to DM relic is found to be $f_{\sigma}\times0.12\times\frac{m_S}{m_\sigma}=6.48\times10^{-3}$, where  $f_\sigma=0.27$. For this choice of parameters, we found that the contribution of $\sigma$ to DM relic density is around $\sim 5\%$. However, its contribution to the $\Delta{N_{\rm eff}}$ is calculated to be 0.858, which is much larger than the current constraints from CMB experiments. In Fig. \ref{fig:rhonuL}, we show the evolution of the energy density of $\nu_L$ originating from $\sigma$ decay, with a red dashed line. The cosmological evolution of SM neutrino energy density is shown with the green dashed line. The neutrino decoupling and the CMB temperatures are shown with the black and gray solid vertical lines, respectively.

We now obtain the correct DM relic contours in the plane of $y_{NS}$ vs $\lambda_{h\sigma}$ for fixed $m_\sigma=500$ GeV, $m_{N_1}=10^4$ GeV, $\delta{M}=2.1\times10^{-4}$ GeV, and $\gamma=0.15-i0.15$ and shown in the \textit{left} panel of Fig. \ref{fig:ynsVSlambda1}. We take two values of DM mass, $m_S=1$ GeV (black line) and $m_S=100$ GeV (gray line). DM relic is obtained considering both the freeze-in ($N_1$ decay) and SuperWIMP ($\sigma$ decay) contributions. The color code represents the percentage of freeze-in component to the total DM relic. As the $\lambda_{h\sigma}$ coupling decreases, the freeze-out abundance of $\sigma$ increases. This leads to an increment in the SuperWIMP component of the DM relic. Now, for the correct relic of DM, the freeze-in part needs to decrease, which implies further smaller values of $y_{NS}$. This behavior can be clearly read from the \textit{left} panel of Fig. \ref{fig:ynsVSlambda1}. As the $\lambda_{h\sigma}$ coupling is increased, the freeze-out abundance of $\sigma$ gets reduced, leading to a smaller SuperWIMP contribution. With very large $\lambda_{h\sigma}$, the freeze-in component becomes almost 100\% (dark blue points), and the relic becomes independent of $y_{NS}$ for a fixed DM mass. For a smaller DM mass, the contour shifts upward. This is because the relic, $\Omega_{\rm DM} h^2\propto m_{S} Y_S$, as $m_S$ decreases, $Y_S$ has to increase, to fix $\Omega_{\rm DM} h^2$. An increment in $y_{NS}$ will lead to an increment in $Y_S$. 

For these same points given in \textit{left} panel of Fig. \ref{fig:ynsVSlambda1}, we calculate the $\Delta{N_{\rm eff}}$ and show as a function of $\lambda_{h\sigma}$ in the \textit{right} panel of Fig. \ref{fig:ynsVSlambda1}. For smaller $\lambda_{h\sigma}$, $\sigma$ decouples with larger abundance. As a result, the correct relic abundance can be obtained from the SuperWIMP component ($\sigma$ decay), leading to a negligible freeze-in component which requires a relatively smaller $y_{NS}$ coupling. When the Yukawa coupling, $y_{NS}$, is small, $\sigma$ decays at a late epoch with a larger $\Delta{N_{\rm eff}}$ as the $\sigma$ abundance is large. On the other hand, an increase of $\lambda_{h\sigma}$ will lead to a decrease in $\sigma$ abundance, leading to a smaller $\Delta{N}_{\rm eff}$ through $\sigma$ decay. This is clearly visible in the \textit{right} panel of Fig. \ref{fig:ynsVSlambda1}. To summarize, a red point implies zero freeze-in component and large $\Delta{N}_{\rm eff}$. Similarly, the blue point represents 100\% freeze-in component with small $\Delta{N}_{\rm eff}$.  The Planck data exclude $\lambda_{h\sigma}\lesssim$0.16 (0.55) for 1 GeV (100 GeV) DM mass. In this parameter space, the contribution of the SuperWIMP component to the DM relic remains $\lesssim 3\%$. The future CMB-HD experiment will be able to exclude $\lambda_{h\sigma}\lesssim$0.55 (2) for 1 GeV (100 GeV) DM mass.

We then fix $m_S=1$ GeV, $m_{N_1}=10^4$ GeV, $\delta{M}=2.1\times10^{-4}$ GeV, and $\gamma=0.15-i0.15$ and calculate the DM relic for two choices of $m_\sigma$ as 100 GeV (black solid line) and 1000 GeV (gray solid line) and shown in Fig. \ref{fig:ynsVSlambda2} in the $y_{NS}-\lambda_{h\sigma}$ plane. The color-code represents the percentage of freeze-in component to the total DM relic, similar to Fig. \ref{fig:ynsVSlambda1}. As the $\lambda_{h\sigma}$ coupling decreases, the freeze-out abundance of $\sigma$ increases, thereby enhancing the SuperWIMP contribution to the DM relic. To obtain the correct relic abundance, the freeze-in component must accordingly decrease, which requires a reduction in the value of $y_{NS}$. This behavior is evident in the \textit{left} panel of Fig.~\ref{fig:ynsVSlambda2}. Conversely, as the $\lambda_{h\sigma}$ coupling increases, the freeze-out abundance of $\sigma$ decreases, leading to a reduced SuperWIMP contribution to the total relic density. For sufficiently large $\lambda_{h\sigma}$, the freeze-in contribution dominates almost entirely (blue points), and the relic abundance becomes independent of $y_{NS}$ for a fixed DM mass. For the same points, we now compute $\Delta N_{\rm eff}$ and present it as a function of $\lambda_{h\sigma}$ in the \textit{right} panel of Fig. \ref{fig:ynsVSlambda2}. Here, the Planck data exclude $\lambda_{h\sigma}\lesssim$0.055 (0.26) for 100 GeV (1000 GeV) $\sigma$ mass. In this parameter space, the contribution of the SuperWIMP component to the DM relic remains $\lesssim 1\%$. The future CMB-HD experiment will be able to exclude $\lambda_{h\sigma}\lesssim$0.19 (0.95) for 100 GeV (1000 GeV) $\sigma$ mass.

Based on the above discussion, we conclude that the current constraints on $\Delta{N}_{\rm eff}$ restrict the SuperWIMP contribution ($\sigma$ decay) to be $\lesssim3\%$ to the total DM relic. Therefore, for $T_{\rm rh}>m_{N_1}$, the DM relic is solely decided by the freeze-in component ($N_1$ decay).

The SuperWIMP contribution remains subdominant once the $\Delta N_{\rm eff}$ constraints are imposed. This behavior can also be understood analytically as discussed below. The SuperWIMP contribution to the dark matter relic abundance scales as
\begin{eqnarray}
    \Omega_{\rm SuperWIMP} h^2 = m_S Y_\sigma \frac{s}{\rho_c},
\end{eqnarray}
where $\rho_c$ is the critical energy density. Each decay of $\sigma$ injects energy into neutrinos of order $E_\nu \simeq m_\sigma/2$, such that the total energy density deposited in $\nu_L$ (non thermal contribution) is $\rho_{\nu_L} \simeq n_\sigma E_\nu \simeq s Y_\sigma E_\nu$. 
The resulting contribution to the relativistic degrees of freedom is therefore
\begin{eqnarray}
    \Delta{N_{\rm eff}}=N_\nu\frac{\rho_{\nu_L}}{\rho_{\nu_L}^{\rm SM}}\simeq \frac{N_\nu sY_\sigma m_\sigma/2}{2\frac{7}{8}\frac{\pi^2}{30}T^4_{\nu_L}}.
\end{eqnarray}
leading to 
\begin{eqnarray}
    \Delta{N_{\rm eff}}\simeq8.3037\times10^{3}\left(\frac{m_\sigma}{m_S}\right)\left(\frac{N_\nu}{3}\right)\left(\frac{\rho_c}{8.1\times10^{-47}{~\rm GeV^{4}}}\right)\left(\frac{2.35\times10^{-13}~{\rm GeV}}{T_{\nu_L}}\right)^4\left(\frac{\Omega_{\rm SuperWIMP}h^2}{0.12}\right).
\end{eqnarray}
Since $m_\sigma > m_S$, a SuperWIMP contribution of $\Omega_{\rm SuperWIMP} h^2 \sim 0.12$ would result in a very large $\Delta N_{\rm eff}$, which is excluded by current CMB observations. This analytic expectation is consistent with our numerical results, where the SuperWIMP fraction remains $\lesssim 3\%$ after imposing all $\Delta N_{\rm eff}$ constraints.

\section{Phenomenology with $T_{\rm EW}<T_{\rm rh}\ll m_{N_1}$: DM production from $h$ decay}\label{sec:lowTrh}


\begin{figure}[ht]
    \centering
    \includegraphics[scale=0.35]{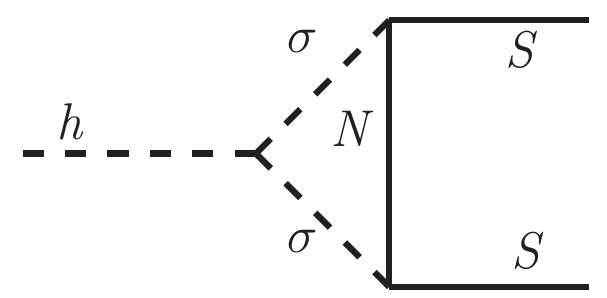}
    \caption{Feynman diagram for DM production from $h$ decay.}
    \label{fig:DMprodhdec}
\end{figure}
\begin{figure*}[tbh]
    \centering
\includegraphics[scale=0.4]{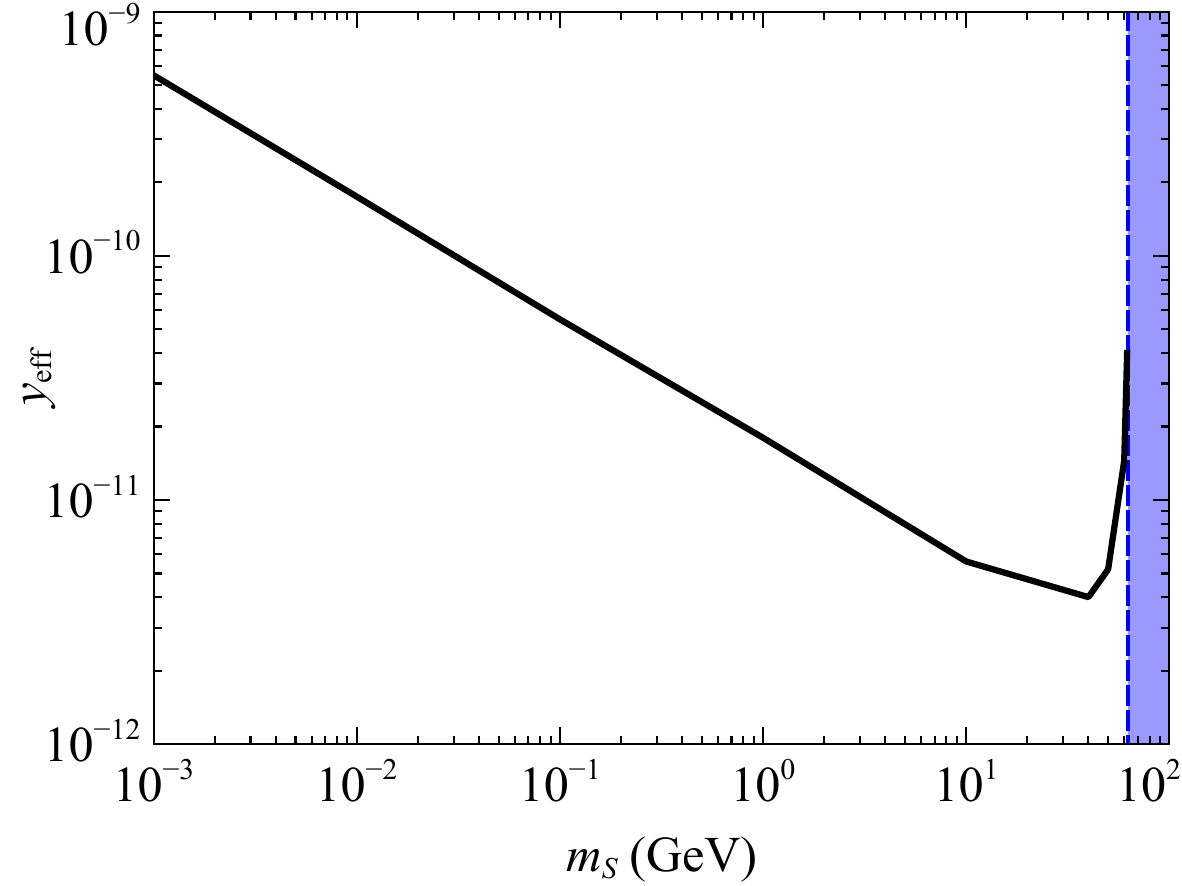}
\includegraphics[scale=0.4]{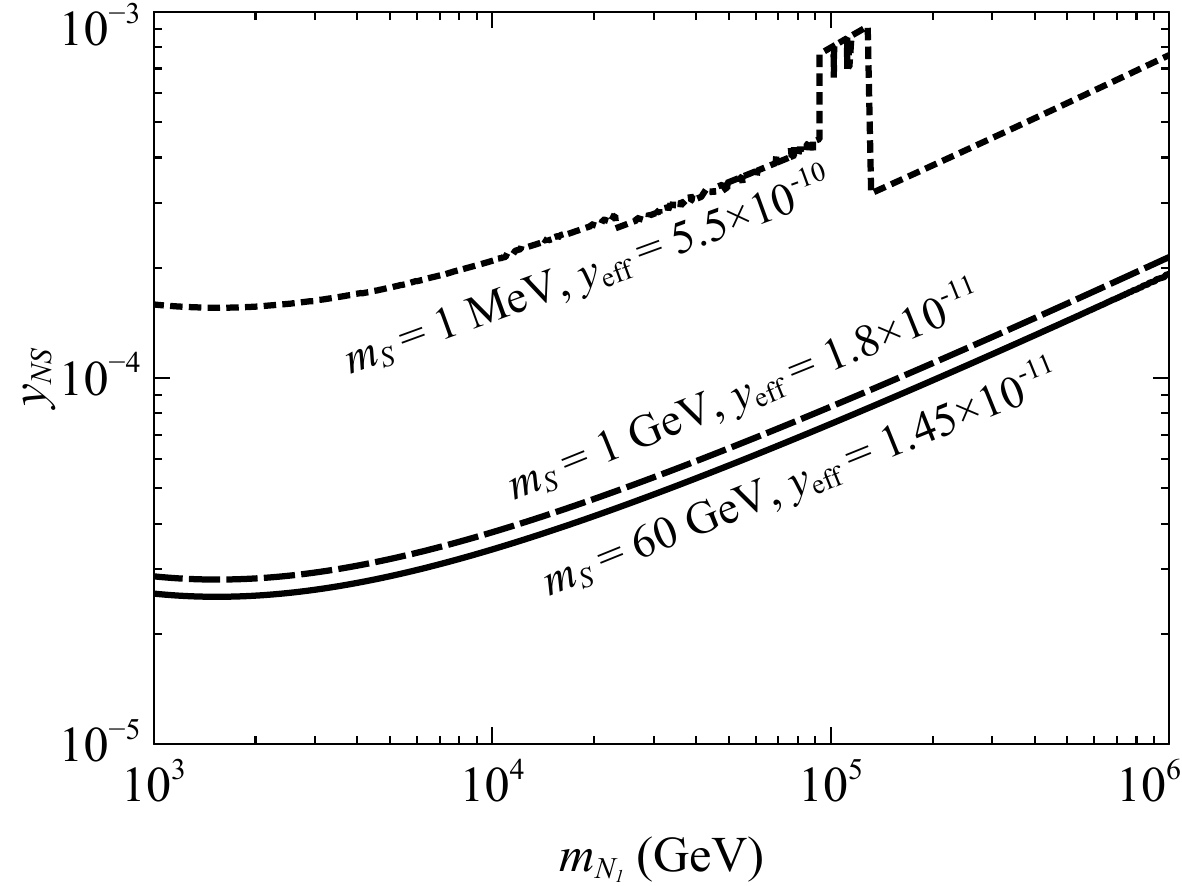}
    \caption{[\textit{Left:}] correct relic contour in the plane $y_{\rm eff}-m_S$. In the blue shaded region, $m_S>m_h/2$. [\textit{Right:}] correct relic contour in the plane $y_{NS}-m_{N_1}$ for $m_S=1$ MeV (dotted), $m_S=1$ GeV (dashed), $m_S=60$ GeV (solid), $m_\sigma=729.1$ GeV, $\lambda_{h\sigma}=6.63$.}
    \label{fig:lowtrh1}
\end{figure*}

We now turn to the possibility of a low reheating scenario with $T_{\rm EW}<T_{\rm rh} \ll m_{N_1}$. In this regime, the reheating temperature is insufficient to produce $N_1$ from the thermal bath. As a result, the tree-level freeze-in channel $N_1 \to S\sigma$ is kinematically inaccessible. The DM candidate $S$ remains out of equilibrium throughout, as its only coupling to the thermal bath is through the loop-suppressed effective vertex $y_{\rm eff} h SS$ (Fig.~\ref{fig:DMprodhdec}), which is proportional to $y_{NS}^2$. This represents a qualitatively distinct production mechanism compared to the high-reheating scenario of Section~\ref{sec:sddmrelic}: the DM relic abundance is determined entirely by SM Higgs decay at one loop, with no dependence on the RHN thermal history, restricting the DM mass to $m_S < m_h/2$. If $T_{\rm rh}>m_\sigma$, then $\sigma$ is also thermalized through $\lambda_{h\sigma}H^\dagger H\sigma^2$. As discussed in the section \ref{sec:foptlargelambda}, if $\lambda_{h\sigma}$ is large such that it can give rise to FOEWPT, then the freeze-out relic of $\sigma$ is small. As a result, the corresponding $S$ relic from $\sigma$ decay is negligibly small. Alternatively, if $\lambda_{h\sigma}$ is small as discussed in section \ref{sec:freezeinandsuperwimp}, then $\sigma$ will freeze out with a large relic. However, in this case, $\sigma$ decay to $S\nu$ will give rise to simultaneous production of $S$ and additional contribution to $N_{\rm eff}$. The decay of $S$ can produce a correct DM relic, but in that case $\Delta{N}_{\rm eff}$ will be unexpectedly large. Thereby ruling out this scenario. Alternatively, if we demand $\Delta{N}_{\rm eff}$ is within the observable order, then the relic of $S$ from $\sigma$ decay is negligibly small. In conclusion $S$ relic can not be realized via $\sigma$ decay. The detectability of this scenario at future GW and CMB experiments remains the same as discussed in Section~\ref{sec:sddmrelic}. Therefore, in this scenario, the DM relic abundance is determined solely by the one–loop decay of the Higgs, $h \to SS$ as shown in Fig. \ref{fig:DMprodhdec}. This restricts the DM mass to $m_S < m_h/2$. The effective vertex can be written as $y_{\rm eff}hSS$, where the effective coupling is given in \ref{app:decayrate}. In this scenario, the Boltzmann equation reduces to
\begin{eqnarray}
\frac{dY_S}{dz}=2\frac{\langle\Gamma_h\rangle}{z\mathcal{H}(T)}Y_h^{\rm eq},\label{eq:BESh}
\end{eqnarray}
where $\langle\Gamma_h\rangle$ is the thermally averaged decay width of Higgs to $SS$ given in \ref{app:decayrate}. We find the correct DM relic contour in the plane of $y_{\rm eff}$ vs $m_S$ and show it with the black solid line in the \textit{left} panel of Fig. \ref{fig:lowtrh1}. The blue shaded region is excluded since $m_S>m_{h}/2$ in that region. From the \textit{left} panel of Fig. \ref{fig:lowtrh1} we see that as the DM mass increases, the $y_{\rm eff}$ coupling decreases to give the correct DM relic. When the DM mass approaches $m_h/2$, the decay width decreases, and hence the abundance. Now, to get the correct relic of DM, $y_{\rm eff}$ has to increase. This behavior is clearly visible in the \textit{left} panel of Fig. \ref{fig:lowtrh1}. $y_{\rm eff}$ is a function of \{$y_{NS},m_{N_1},m_{\sigma},\lambda_{h\sigma},m_S$\}. We then take three sets of \{$m_S,y_{\rm eff}$\} values from the \textit{left} panel of Fig. \ref{fig:lowtrh1} and show the allowed values of $y_{NS}$ and $m_{N_1}$ by fixing $m_{\sigma}=729.1$ GeV and $\lambda_{h\sigma}=6.63$. This choice of parameters leads to FOEWPT, as shown with a cyan star in Fig. \ref{fig:gwamp}.

\section{Conclusion}\label{sec:concl}

In this study, we have explored the realization of freeze-in dark matter within a minimal extension of the type-I seesaw framework, where the remnant lepton-parity symmetry naturally plays the role of a dark parity. By introducing a lepton parity-even singlet fermion $S$ and a lepton parity-odd scalar $\sigma$, the setup naturally accommodates a stable dark matter candidate ($S$). In our setup, the DM remains out of equilibrium throughout the evolution. Therefore, its relic is determined by the thermal history of the early Universe. For reheating temperatures, $T_{\rm rh}>m_{N_1}$, where $m_{N_1}$ is the mass of the lightest RHN ($N_1$), the dominant contribution to the DM relic density originates from right-handed neutrino decays ($N_1\rightarrow LH$), while for $T_{\rm EW}<T_{\rm rh}\ll m_{N_1}$,  it can be produced via Higgs decay ($h\rightarrow SS$). Direct and indirect detection constraints on DM are naturally avoided due to the extremely small Yukawa couplings. However, this scenario can leave observable imprints through gravitational waves and as well as CMB experiments. When the $h-\sigma$ coupling is larger (typically $\lambda_{h\sigma}>1$), then the EWPT can be first-order and gives rise to stochastic gravitational waves which can be detected at different GW experiments like DECIGO, BBO, etc. On the other hand if $h-\sigma$ coupling is small (typically $\lambda_{h\sigma}\ll1$), then the late decay of $\sigma$ can give rise observable $\Delta{N}_{\rm eff}$.

The decay of $N$ not only gives rise to the DM relic but also generates lepton asymmetry in the early Universe through $N\rightarrow LH$ decay. To realize low-scale leptogenesis, we consider nearly degenerate right-handed neutrinos so that the \textit{CP} asymmetry can be resonantly enhanced. The lepton asymmetry thus produced is subsequently converted into the observed baryon asymmetry through electroweak sphaleron processes.

In conclusion, our model provides a scenario connecting dark matter production, leptogenesis, $\Delta{N}_{\rm eff}$, and electroweak symmetry breaking dynamics in a unified and testable framework. This could be possible, in comparison to our previous analysis \cite{Ma:2024dhk, Ma:2025bjf} due to the existence of additional $NS\sigma$ coupling in our model.

\section*{Acknowledgements}
P.K.P. acknowledges the Ministry of Education, Government of India, for providing financial support for his research via the Prime Minister’s Research Fellowship (PMRF) scheme. P.K.P. would like to thank Santu Kumar Manna for his assistance with {\tt CosmoTransitions}.

\appendix
\section{Flavored-resonant leptogenesis}\label{app:lepto}

The Boltzmann equation for $N_1$ abundance is already given in Eq. \ref{eq:N1}. The Boltzmann equations driving the abundance of $N_2$, and asymmetries in different flavors are given as
\begin{eqnarray}
    \frac{dY_{N_2}}{dz}=-(D_2+S^t_2+S^s_2)(Y_{N_2}-Y^{\rm eq}_{N_2}),\label{eq:N2}
\end{eqnarray}
\begin{eqnarray}
    \frac{dY_{\Delta e}}{dz}&=&\epsilon^1_{ee}D_1(Y_{N_1}-Y^{\rm eq}_{N_1})+\epsilon^2_{ee}D_2(Y_{N_2}-Y^{\rm eq}_{N_2})-\left(W^D_1+W^t_1+\frac{Y_{N_1}}{Y^{\rm eq}_{N_1}}W^s_1\right) \left(\frac{|y_{e1}|^2}{(y^\dagger y)_{11}}\right) Y_{\Delta e}\nonumber\\&&-\left(W^D_2+W^t_2+\frac{Y_{N_2}}{Y^{\rm eq}_{N_2}}W^s_2\right)\left(\frac{|y_{e2}|^2}{(y^\dagger y)_{22}}\right)Y_{\Delta e}
    -W_{\Delta L2}Y_{\Delta e},
    \label{eq:Nee}
\end{eqnarray}
\begin{eqnarray}
    \frac{dY_{\Delta \mu}}{dz}&=&\epsilon^1_{ \mu \mu}D_1(Y_{N_1}-Y^{\rm eq}_{N_1})+\epsilon^2_{ \mu \mu}D_2(Y_{N_2}-Y^{\rm eq}_{N_2})-\left(W^D_1+W^t_1+\frac{Y_{N_1}}{Y^{\rm eq}_{N_1}}W^s_1\right) \left(\frac{|y_{ \mu1}|^2}{(y^\dagger y)_{11}}\right) Y_{\Delta \mu}\nonumber\\&&-\left(W^D_2+W^t_2+\frac{Y_{N_2}}{Y^{\rm eq}_{N_2}}W^s_2\right)\left(\frac{|y_{ \mu2}|^2}{(y^\dagger y)_{22}}\right)Y_{\Delta \mu}
    -W_{\Delta L2}Y_{\Delta \mu},\label{eq:Nmm}
\end{eqnarray}
\begin{eqnarray}
    \frac{dY_{\Delta \tau}}{dz}&=&\epsilon^1_{\tau\tau}D_1(Y_{N_1}-Y^{\rm eq}_{N_1})+\epsilon^2_{\tau\tau}D_2(Y_{N_2}-Y^{\rm eq}_{N_2})-\left(W^D_1+W^t_1+\frac{Y_{N_1}}{Y^{\rm eq}_{N_1}}W^s_1\right) \left(\frac{|y_{\tau1}|^2}{(y^\dagger y)_{11}}\right) Y_{\Delta\tau}\nonumber\\&&-\left(W^D_2+W^t_2+\frac{Y_{N_2}}{Y^{\rm eq}_{N_2}}W^s_2\right)\left(\frac{|y_{ \tau2}|^2}{(y^\dagger y)_{22}}\right)Y_{\Delta\tau}
    -W_{\Delta L2}Y_{\Delta \tau},\label{eq:Ntt}
\end{eqnarray}
The abundance, $Y_i$ is defined as $Y_i=n_i/s$. $D_{i}=\langle\Gamma_{N_i\rightarrow LH}\rangle/\mathcal{H}z$, where $\langle\Gamma_{N_i\rightarrow LH}\rangle$ is the thermally averaged decay width for $N_i\rightarrow LH$. $S^{s,t}_i=\Gamma^{s,t}_i/\mathcal{H}z$ are the scattering terms which include both Yukawa and gauge boson contributions, $W^D_i$ corresponds to the washout term due to inverse decay, $W^{s,t}_i$ represent the washout due to $\Delta L=1$ scatterings considering both Yukawa coupling driven processes and gauge boson involved processes and $W_{\Delta L2}$ represents the $\Delta L=2$ washout. We follow the notations of \cite{Buchmuller:2004nz}.

\begin{figure}[ht]
    \centering
    \includegraphics[scale=0.42]{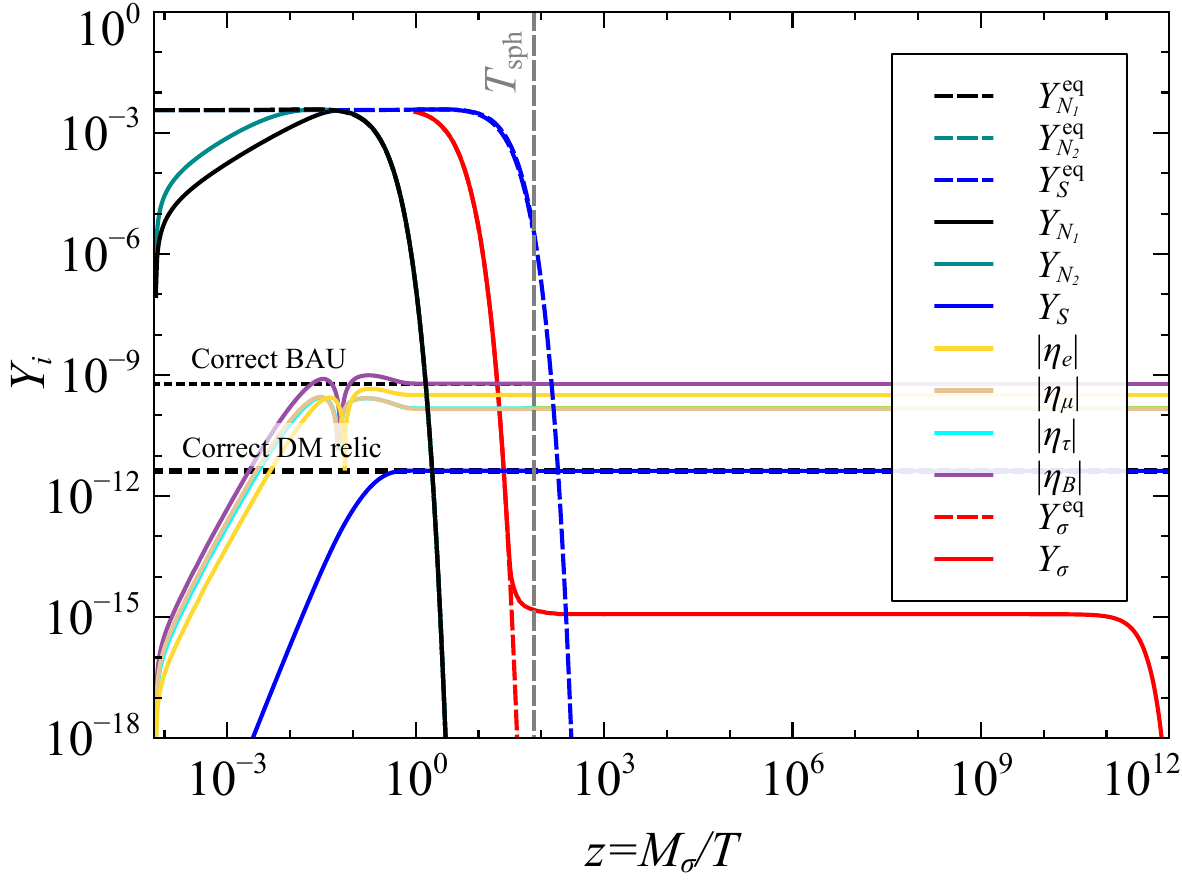}
    \caption{Cosmological evolution of the abundances of $N_1,N_2,\sigma$, $S$, and lepton asymmetries. The corresponding equilibrium abundances are shown with the dashed line. The parameters are fixed at \{$m_{N_1}=10^{4}$ GeV, $m_\sigma=729.1$ GeV, $\lambda_{h\sigma}=6.63$, $m_S=100$ GeV, $y_{NS}=1.56\times10^{-11}$, $\delta{M}=2.1\times10^{-4}$ GeV,$\gamma=0.15-i0.15$ \}. Here $\eta_i=7.04\frac{C}{f}Y_{\Delta i}$. The vertical gray dashed line represents the sphaleron transition temperature.}
    \label{fig:lepto}
\end{figure}

In Fig. \ref{fig:lepto}, we show the evolution of the asymmetries as well as the abundances of RHNs and DM. The parameters are fixed at the same value as in Fig. \ref{fig:bp1N}.
The vertical gray dashed line represents the sphaleron transition temperature. The observed baryon asymmetry and DM relic values are indicated with black dashed lines.

\section{Finite-Temperature Effective Potential}\label{app:fopt}
The finite temperature effective potential can be written as
\begin{eqnarray}
    V_{\rm eff}=V_0(h)+V_{\rm cw}(h)+V_{\rm ct}(h)+V_T(h,T)+V_{\rm daisy}(h,T),
\end{eqnarray}
where the tree-level potential is given as
\begin{eqnarray}
    V_0(h)=-\frac{\mu_h^2}{2}h^2+\frac{\lambda_h}{4}h^4.
\end{eqnarray}
The one-loop Coleman-Weinberg zero temperature contribution in $\overline{\rm MS}$ renormalization scheme is given as
\begin{eqnarray}
    V_{\rm cw}(h)=\sum_{i} (-1)^{F_i}\frac{n_i^2}{64\pi^2}m^2_i(h)\left( \log\left( \frac{m_i^2(h)}{m_i^2(v_h)} \right)-c_i \right),\nonumber\\
\end{eqnarray}
where $F_i$ denotes fermion number, which is 1 for fermions and 0 for bosons, $c_i=3/2$ for fermions and scalars, and $c_i=5/6$ for vector bosons, $m_i(h)$ are the fields dependent masses given as
\begin{eqnarray}
   m_h^2(h)=-\mu_h^2+3\lambda_h h^2,~n_h=1,\nonumber\\
   m_{GB}^2(h)=-\mu_h^2+\lambda_h h^2,~n_{GB}=1,\nonumber\\
   m_\sigma^2(h)=\mu_\sigma^2+\lambda_{h\sigma} h^2,~n_\sigma=1,\nonumber\\
   m_t^2(h)=\frac{y_t^2}{2} h^2,~n_t=12,\nonumber\\
   m_W^2(h)=\frac{g^2}{4} h^2,~n_W=6,\nonumber\\
   m_Z^2(h)=\frac{g^2+g^{\prime2}}{4} h^2,~n_Z=3.
\end{eqnarray}
The counter term is
\begin{eqnarray}
    V_{\rm ct}(h)=-\frac{\delta\mu_h^2}{2}h^2+\frac{\delta\lambda_h}{4}h^4,
\end{eqnarray}
which is derived by solving
\begin{eqnarray}
    \frac{\partial(V_{\rm cw}+V_{\rm ct})}{\partial h}|_{h=v_h}=0, \frac{\partial^2(V_{\rm cw}+V_{\rm ct})}{\partial h^2}|_{h=v_h}=0.
\end{eqnarray}
The finite temperature contribution to the effective potential is given as
\begin{eqnarray}
    V_T(h,T)=\frac{T^4}{2\pi^2}\left[ \sum_{i\in B} n_i J_B\left(\frac{m_i(h)}{T}\right)- \sum_{j\in F} n_j J_F\left(\frac{m_j(h)}{T}\right) \right],\nonumber\\
\end{eqnarray}
where
\begin{eqnarray}
J_{B/F}(y)=\int_0^{\infty} x^2\log(1\mp e^{-\sqrt{x^2+y^2}})
\end{eqnarray}
The daisy contribution is
\begin{eqnarray}
    V_{\rm daisy}(h,T)=\sum_i\frac{T}{12\pi}n_i\left[ m_i^3(h)-\left(m_i^2(h) +\Pi_i(T) \right)^{3/2} \right],\nonumber\\
\end{eqnarray}
where $n_i$ denotes all d.o.f. for scalars and only longitudinal d.o.f. for vector bosons and $\Pi_i$s are given as
\begin{eqnarray}
    \Pi_h(T)&=&T^2\left( \frac{1}{2}\lambda_h+\frac{1}{4}y_t^2+ \frac{3}{16}g^2 +\frac{1}{16}g^{\prime 2} +\frac{1}{12}\lambda_{h\sigma} \right),\nonumber\\
    \Pi_{GB}(T)&=&T^2\left( \frac{1}{2}\lambda_h+\frac{1}{4}y_t^2+ \frac{3}{16}g^2 +\frac{1}{16}g^{\prime 2} +\frac{1}{12}\lambda_{h\sigma} \right),\nonumber\\\Pi_\sigma(T)&=&T^2\left(\frac{1}{3}\lambda_{h\sigma} \right),\nonumber\\  
    \Pi_W(T)&=&T^2\left( \frac{11}{6}g^2 \right),\nonumber\\ 
    \Pi_Z(T)&=&\frac{1}{2}\left[ (g^2+g^{\prime2})(\frac{11}{6}T^2-\frac{1}{4}h^2) \right.\left.+ \sqrt{ (g^2-g^{\prime2})^2\left(\frac{11}{6}T^2+\frac{1}{4}h^2\right)^2+\frac{1}{4}g^2g^{\prime2}h^4 }  \right],\nonumber\\ 
    \Pi_A(T)&=&\frac{1}{2}\left[ (g^2+g^{\prime2})(\frac{11}{6}T^2+\frac{1}{4}h^2) \right.\left.- \sqrt{ (g^2-g^{\prime2})^2\left(\frac{11}{6}T^2+\frac{1}{4}h^2\right)^2+\frac{1}{4}g^2g^{\prime2}h^4 }  \right],\nonumber\\ 
\end{eqnarray}

\section{Decay widths}\label{app:decayrate}

\begin{figure}[ht]
    \centering
    \includegraphics[scale=0.3]{sigmatoSnu.pdf}
    \includegraphics[scale=0.3]{htoSS.pdf}
    \includegraphics[scale=0.35]{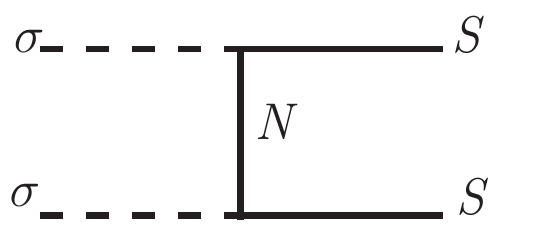}
    \includegraphics[scale=0.3]{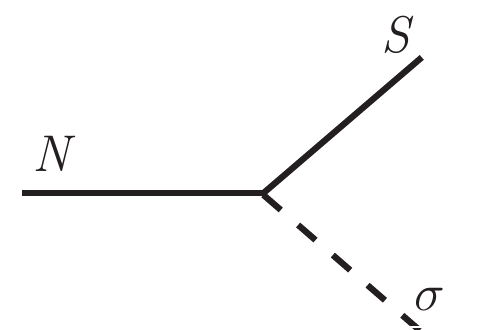}
    \includegraphics[scale=0.3]{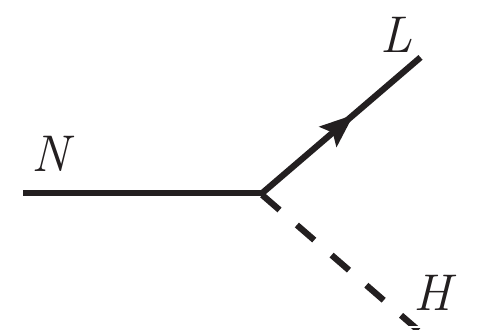}
    \caption{DM production from $\sigma$ decay, $h$ decay, $\sigma\sigma$ annihilation, $N$ decay. DM can thermalized via $\sigma\sigma\rightarrow SS$ channel and lepton asymmetry production from N decay. The order of the plots is from left to right.}
    \label{fig:DMprod}
\end{figure}
The different decay widths for the processes shown in Fig. \ref{fig:DMprod} are given as
\begin{eqnarray}
    \Gamma_{N_1\rightarrow LH}=\frac{(y^\dagger y)_{11}}{8\pi}m_{N_1}
\end{eqnarray}
\begin{eqnarray}
    \Gamma_{N\rightarrow \sigma S}=\frac{y_{NS}^2}{16\pi m_{N_1}^3}\left( (m_{N_1}+m_S)^2-m_\sigma^2 \right)\sqrt{m_\sigma^4+(m_{N_1}^2-m_S^2)^2-2m_\sigma^2(m_{N_1}^2+m_S^2)}
\end{eqnarray}
\begin{eqnarray}
    \Gamma_{\sigma\rightarrow S\nu}=\frac{m_\sigma}{32\pi}\frac{y_{NS}^2y^2v_h^2}{m_{N_1}^2}\left( 1-\frac{m_s^2}{m_\sigma^2} \right)^2
\end{eqnarray}
\begin{eqnarray}
    \Gamma_{h\rightarrow SS}=\frac{y_{\rm eff}^2}{16\pi}m_h\left( 1-\frac{4m_S^2}{m_h^2} \right)^{3/2}, 
\end{eqnarray}
where 
\begin{eqnarray}
y_{\rm eff}=\frac{y_{NS}^2\lambda_{h\sigma}m_{N_1} v_h \mathcal{C}_0[m_h^2,m_S^2,m_S^2,m_\sigma,m_\sigma,m_{N_1}]}{4\sqrt{2}\pi^2},
\end{eqnarray}
where $\mathcal{C}_0[m_h^2,m_S^2,m_S^2,m_\sigma,m_\sigma,m_{N_1}]$ is the scalar Passarino-Veltman three-point function.

The thermally averaged decay widths are given as
\begin{eqnarray}
    \langle\Gamma_{\sigma}\rangle=\Gamma_{\sigma}\frac{K_1(m_\sigma/T)}{K_2(m_\sigma/T)};~~\langle\Gamma_{h}\rangle=\Gamma_{h}\frac{K_1(m_h/T)}{K_2(m_h/T)}.
\end{eqnarray}

\section{Thermally averaged cross-sections}\label{app:crossec}
We calculate the cross-sections using {\tt CalcHEP} \cite{Belyaev:2012qa}. The thermally averaged cross-sections are then given as

\begin{eqnarray}
    \langle\sigma v\rangle_{ab\rightarrow cd}=\frac{T}{32\pi^4}\frac{g_ag_b}{n_a^{\rm eq}n_b^{\rm eq}}\int_{{\rm max}[(m_a+m_b)^2,(m_c+m_d)^2]}^{\infty}ds\sigma(s)\frac{\left[(s-m^2_a-m^2_b)^2-4m_a^2m_b^2\right]}{\sqrt{s}}K_1\left(\frac{\sqrt{s}}{T}\right)
\end{eqnarray}

\section{Direct detection of dark matter}\label{sec:dd}

\begin{figure}[ht]
    \centering
    \includegraphics[scale=0.4]{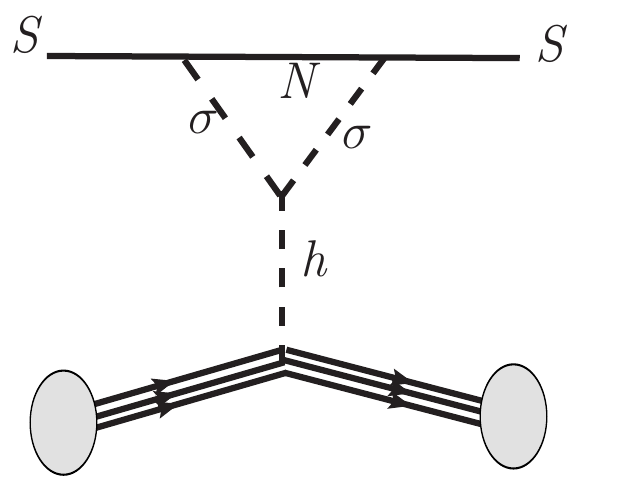}
    \caption{Feynman Diagram for spin-independent DM-nucleon scattering at one loop.}
    \label{fig:ddFD}
\end{figure}

The DM can scatter off with the target nucleus via a one-loop diagram as shown in Fig. \ref{fig:ddFD}.
The effective coupling between the DM and the SM Higgs can be written from Fig. \ref{fig:ddFD} as
\begin{eqnarray}
    y^{\rm eff}_{hSS}=\frac{y_{NS}^2\lambda_{h\sigma}v_hm_{N_1}}{8\sqrt{2}\pi^2m_S}\left[ \log\left( \frac{m_{N_1}^2}{m_\sigma^2} \right) \right. \left.-\frac{2\mathcal{A}}{\sqrt{\mathcal{B}\mathcal{C}}} \log\left( \frac{\mathcal{A}-2m_{N_1}^2+\sqrt{\mathcal{B}\mathcal{C}}}{2m_{N_1}m_\sigma} \right) \right],
\end{eqnarray}
where $\mathcal{A}=m_{N_1}^2+m_S^2-m_\sigma^2$, $\mathcal{B}=(m_{N_1}+m_\sigma)^2-m_S^2$, $\mathcal{C}=(m_{N_1}-m_\sigma)^2-m_S^2$.

The spin-independent cross-section is given as
\begin{eqnarray}
    \sigma^{\rm SI}_{\rm DM-N}=\frac{\mu^2}{\pi A^2}\mathcal{M}^2
\end{eqnarray}
were $\mu$ is the reduced mass of the DM-nucleon system, $A$ is the mass number of the target nucleus, and $\mathcal{M}$ is the amplitude of the diagram shown in Fig. \ref{fig:ddFD} and is given as
\begin{eqnarray}
    \mathcal{M}=Z*f_p+(A-Z)*f_n,
\end{eqnarray}
with $Z$ being the atomic number and interaction strengths between the DM and proton, neutron are given as
\begin{eqnarray}
    f_p=\sum_{q\in u,d,s}f_p^q\alpha^q\frac{m_p}{m_q}+\frac{2}{27}\sum_{Q\in c,b,t}f_p^Q\alpha^Q\frac{m_p}{m_Q},\nonumber\\
    f_n=\sum_{q\in u,d,s}f_n^q\alpha^q\frac{m_n}{m_q}+\frac{2}{27}\sum_{Q\in c,b,t}f_n^Q\alpha^Q\frac{m_n}{m_Q},
\end{eqnarray}
where
\begin{eqnarray}
    \alpha^q=y^{\rm eff}_{hSS}\frac{m_q}{v_h}\frac{1}{m_h^2},~\alpha^Q=y^{\rm eff}_{hSS}\frac{m_Q}{v_h}\frac{1}{m_h^2}.
\end{eqnarray}

Now we compute the direct detection cross-section taking two point from the \textit{left} panel of Fig. \ref{fig:lowtrh1} as \{$m_S=1 ~{\rm MeV},y_{\rm eff}=5.5\times10^{-10}$\}, and \{$m_S=62 ~{\rm GeV},y_{\rm eff}=4\times10^{-11}$\}. The cross-sections for these two points are $2.78\times10^{-67}~\rm cm^2$ and $1.26\times10^{-63}~\rm cm^2$, respectively. These are well below the current constraints on the spin-independent cross-section from the LZ \cite{LZ:2024zvo} experiment or the future sensitivity of DARWIN \cite{DARWIN:2016hyl}.


\end{document}